\documentclass[aps,pra,amsfonts,amssymb,amsmath,showpacs,
floatfix,nofootinbib,citesort]{revtex4-2}
\usepackage{mathrsfs}
\usepackage{amsfonts}
\usepackage{amstext}
\usepackage{amsmath}
\usepackage{amssymb}
\usepackage{bm}
\usepackage{CJK}
\usepackage{bbm}
\usepackage[dvips]{graphicx}
\def\qed{\leavevmode\unskip\penalty9999 \hbox{}\nobreak\hfill
	\quad\hbox{\leavevmode  \hbox to.77778em{%
			\hfil\vrule   \vbox to.675em%
			{\hrule width.6em\vfil\hrule}\vrule\hfil}}
	\par\vskip3pt}

\usepackage{amssymb}
\usepackage{graphicx}
\usepackage{graphics}
\usepackage{amsmath}
\usepackage{amsthm}
\usepackage{color}
\usepackage{dsfont}
\usepackage{textcomp}
\usepackage{subfig}
\usepackage{threeparttable}
\usepackage{booktabs}
\usepackage{diagbox}
\usepackage{float}
\definecolor{darkred}  {rgb}{0.5,0,0}
\definecolor{darkblue} {rgb}{0,0,0.5}
\definecolor{darkgreen}{rgb}{0,0.5,0}
\usepackage{hyperref}
\hypersetup{
	pdftitle = {QRT Proposal},
	pdfauthor = {},
	colorlinks = true,
	urlcolor  = blue,         
	linkcolor = red,     
	citecolor = blue,    
	filecolor = darkred       
}
\usepackage{mathtools}
\def\ra{\rangle}
\def\la{\langle}

\newcommand{\bea}{\begin{eqnarray}}
	\newcommand{\eea}{\end{eqnarray}}
\newcommand{\be}{\begin{equation}}
	\newcommand{\ee}{\end{equation}}
\newcommand{\ba}{\begin{equation}\begin{aligned}}
		\newcommand{\ea}{\end{aligned}\end{equation}}

\newcommand{\beq}{\begin{eqnarray}}
\newcommand{\eeq}{\end{eqnarray}}
\newcommand{\nn}{\nonumber}

\newcommand{\beax}{\begin{eqnarray*}}
	\newcommand{\eeax}{\end{eqnarray*}}
\newcommand{\bex}{\begin{equation*}}
	\newcommand{\eex}{\end{equation*}}

\theoremstyle{remark}

\def\be{\begin{equation}}
	\def\ee{\end{equation}}

\newcommand{\tr}{{\rm Tr}}

\begin{document}
	

\preprint{APS/123-QED}
\begin{CJK*}{GB}{gbsn}
	\title{Quantum Dynamics of Interacting oscillators in a thermal medium: A novel scheme\\}
		

		\author{R. Moradi}
		\author{F. Kheirandish}
		\email{f.kheirandish@uok.ac.ir}
		
		\affiliation{Department of Physics, Faculty of Science, University of Kurdistan, P O Box 66177-15175, Sanandaj, Iran}


\begin{abstract}
We develop an exact analytical framework for studying the nonequilibrium dynamics of interacting quantum harmonic oscillators coupled to a modelled thermal reservoir and driven by external classical fields. The total Hamiltonian is diagonalized via successive Bogoliubov transformations, enabling a nonperturbative treatment of dissipation and driving. For two coupled oscillators with an external source applied to the first, we derive explicit energy expressions and analyze their dependence on coupling and bath parameters. The phase-space structure is characterized through Husimi $Q$-functions for initial separable coherent and number states.  We obtain the reduced density matrix elements in the number-state basis and demonstrate that, in the absence of driving and at zero temperature, the reduced density matrix of the main system satisfies the Lindblad master equation. A generating function for the Husimi function is introduced to facilitate computation of higher-order correlations. The analysis is then generalized to $n$ interacting oscillators, with analytical energy expressions provided and the case $n=3$ examined in detail. Our results establish a versatile and exact framework for driven-dissipative quantum systems, with potential applications in quantum thermodynamics, open quantum systems, and many-body physics.
\end{abstract}
		
		\maketitle
	\end{CJK*}

\section{Introduction}\label{Sec1}
\noindent
Many physical systems such as atoms in external fields \cite{Zimmerman1979}, coupled quantum dots \cite{Bayer2001}, and cavity optomechanics \cite{Kippenberg2008}, can be described generically using coupled mechanical oscillators as a model system. The study of coupled quantum harmonic oscillators in the presence of dissipation \cite{Zhao2003} represents a cornerstone of modern quantum thermodynamics, quantum information science, and condensed matter physics. While isolated interacting oscillators have been extensively characterized, real-world quantum systems inevitably couple to their environments, leading to decoherence, dissipation, and perhaps counter intuitively, new emergent phenomena such as dissipative quantum phase transitions and environment-induced synchronization.,,

Recent years have witnessed a surge of interest in the rich interplay between interactions, dissipation, and quantum correlations, driven by advances in experimental platforms including superconducting circuits \cite{Zagoskin2026}, trapped ions \cite{Milburn2025,Garraway2025}, optomechanical systems \cite{Milburn2015}, and photonic cavities \cite{Castelletto2025}. These platforms enable unprecedented control over both coherent coupling strengths and engineered dissipation channels, opening new avenues for quantum state engineering and non-equilibrium many-body physics.

The dissipative quantum Rabi model, describing a two-level system coupled to a damped harmonic oscillator, has long served as a paradigmatic open quantum system. A recent study \cite{Kang2026} reveals that oscillator dephasing constitutes a relevant perturbation that fundamentally alters the nature of the steady-state phase transition. Specifically, they demonstrate that dephasing drives the system from a Gaussian to a non-Gaussian phase transition, accompanied by an intriguing cascade of instabilities. This finding challenges conventional Gaussian approximations and underscores the importance of properly treating all dissipative channels in interacting oscillator-bath systems.

The coordinated behavior of coupled oscillators, familiar from classical metronome synchronization, finds a powerful quantum analog in dissipatively coupled systems. In \cite{Li2026}, authors engineered a two-qubit system interacting via a shared thermal environment, deriving an effective master equation containing a two-body dissipator. By quenching this dissipator, they demonstrated controlled transitions between in-phase and anti-phase synchronization states, a phenomenon remarkably robust against external noise. Their experimental validation using superconducting circuits establishes dissipative coupling as a resource rather than a nuisance. Extending this theme, authors in \cite{Dai2025} achieved universal tuning of quantum synchronization in spin oscillator networks by continuously varying interaction anisotropy. They uncovered the existence of a complete synchronization blockade under purely directional coupling, an exclusively quantum effect that suppresses coordinated oscillations entirely, providing a general framework applicable from few-body to many-body systems.

In \cite{Mivehvar2026}, it was shown that a system comprising Landau levels of a charge-neutral particle in a synthetic gauge field coupled to an optical cavity can be exactly mapped onto two highly nonlinearly coupled quantum harmonic oscillators with dissipation. This mapping unveils the formation of Landau polaritons, hybrid light-matter states exhibiting nonzero entanglement and quadrature squeezing, and reveals multiple steady states with distinct physical properties depending on initial conditions. The work provides an analytically tractable playground for exploring non-equilibrium dynamics in driven-dissipative settings.

In \cite{Luo2025}, authors investigated optimal entanglement transfer in coupled oscillator chains embedded in generic non-Markovian environments. Employing the Krotov optimization algorithm adapted for time-dependent dissipative terms, they demonstrated high-fidelity transfer achievable simply by tuning oscillator frequencies while keeping coupling strengths constant. Remarkably, quantum memory effects arising from non-Markovian environmental coupling aid entanglement transfer, outperforming the memoryless (Markovian) case, a counterintuitive result with direct implications for quantum repeater design.

A fundamental methodological advance came from a study of two coupled quantum oscillators in a common Lorentzian environment, solved exactly without Born or Markov approximations \cite{Wu2026}. The authors showed that controlled frequency detuning suppresses decoherence and non-Markovian revivals, significantly reducing system-bath correlations. These results provide practical design rules for protecting quantum coherence through dynamical control. Complementing this, a recent study of three coupled oscillators connected to two independent thermal baths at different temperatures compared exact dynamics with global and local master equation approximations \cite{Babakan2026}. The authors identified a temperature-dependent critical inter-oscillator coupling strength determining which approximation scheme better reproduces exact evolution, a crucial benchmark for modeling larger dissipative networks.

Beyond the linear regime, a dissipative charged anharmonic oscillator in a magnetic field analyzed in \cite{Mandal2025}. Using perturbative techniques within a non-Markovian master equation framework, they demonstrated that anharmonicity enhances decoherence, a deconfining effect, and increases von Neumann entropy. The oscillatory nature of the heating function reveals information backflow, a hallmark of non-Markovian memory effects.

In our previous works \cite{Kheir1_2025,Kheir2_2025,Kheir2026}, we introduced a scheme for investigating the quantum dynamics of a primary system interacting with a thermal bath. The central idea is as follows: a primary system with Hamiltonian $\hat{H}_S$ is placed in contact with a thermal reservoir at temperature $T$, modeled by an identical system with Hamiltonian $\hat{H}_B$. When the primary system is initially at a higher temperature, it transfers heat to the bath. Owing to the monotonically decaying coupling function, this energy flow is effectively unidirectional, with negligible back flow to the system. Conversely, when the primary system is colder than the bath, the symmetry of the composite system reverses the heat current, driving it from the bath into the primary system. We apply this scheme to investigate the quantum dynamics of two interacting oscillators in a thermal bath.

The layout and the results of the paper are as follows:
In Sec.~\ref{Sec2}, we investigate the dynamics of two interacting quantum harmonic oscillators coupled to a modelled thermal bath in the presence of an external source applied to the first oscillator, within the framework of a novel scheme. In this section, the Hamiltonian of the total system is diagonalized through successive Bogoliubov transformations, and explicit expressions for the energy of the oscillators are obtained and depicted in Fig.~(\ref{Fig2}). In Sec.~\ref{Sec3}, the Husimi $Q$-functions for both oscillators, along with the corresponding reduced $Q$-functions, are explicitly derived for initially prepared states corresponding to separable coherent and number states. The positions of the maxima of the Husimi function are illustrated in Fig.~(\ref{Fig3}) for various parameter values at zero temperature. Additionally, an explicit expression for the components of the reduced density matrix in the number-state basis is obtained. In the subsequent subsections, it is demonstrated that, in the absence of an external driving force and at zero temperature, the reduced density matrix of the main system satisfies the Lindblad master equation. Finally, a generating function for the Q-function is introduced. In Sec.~\ref{Sec4}, the analysis is generalized to a system of $n$ interacting oscillators embedded in a thermal medium and driven by an external classical field. Explicit analytical expressions for the energy of each oscillator are derived, and the specific case of three coupled oscillators ($n=3$) is examined in detail.
\section{Interacting oscillators in a thermal medium}\label{Sec2}
\begin{figure}[h]
\centering
\includegraphics[scale=0.2]{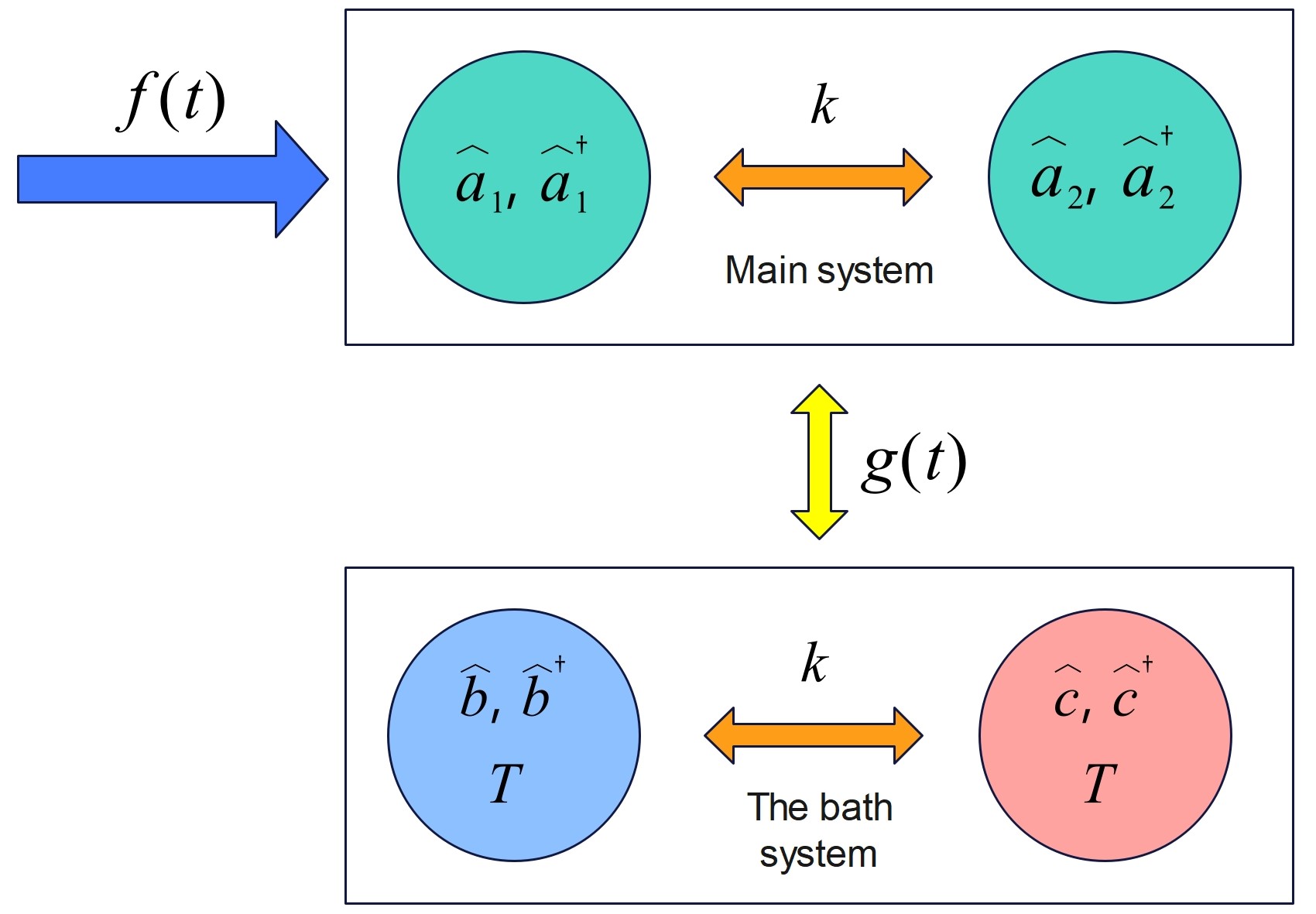}
\caption{(Color online) Two interacting oscillators as the main system, interact through the time-dependent coupling function $g(t)$ with the bath system which is a copy of the main system. The bath system is initially prepared in a thermal state with inverse temperature $\beta=1/\kappa_B T$, and mean excitation numbers $\bar{n}_b=\bar{n}_c=\bar{n}=(e^{\beta\hbar\omega_0}-1)^{-1}$. A deriving classical force with a time-dependent amplitude $f(t)$ and frequency $\omega_L$ is applied to the first oscillator.}\label{Fig1}
\end{figure}

\begin{figure}[h]
\centering
\includegraphics[scale=0.8]{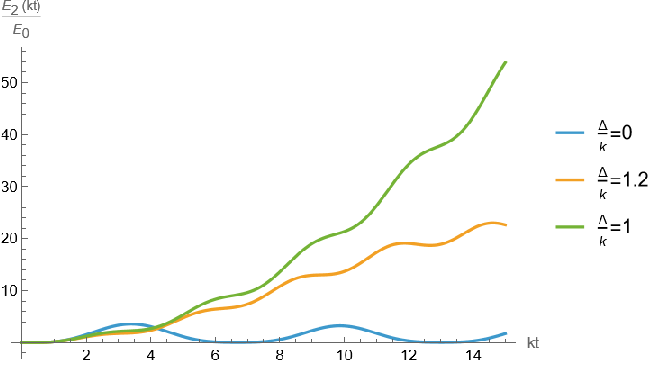}
\caption{(Color online) The scaled energy of the right oscillator for different values of the scaled detuning $\frac{\triangle}{k}=\frac{\omega_0-\omega_L}{k}=0,\,1,\,1.2.$ The amplitude of the external source is assumed constant $f(t)=F$, and the energy is rescaled by $F^2/k^2$. For the value $\frac{\triangle}{k}=1$, the energy increases without bound.}\label{Fig2}
\end{figure}

\noindent
In this section, we investigate the dynamics of two interacting quantum harmonic oscillators in a thermal bath under the influence of a classical driving force with a time-dependent amplitude $f(t)$, and frequency $\omega_L$, applied to the first oscillator. The bath system is by assumption a copy of the main system initially prepared in a thermal state with inverse temperature $\beta=1/\kappa_B T$, see Fig.~(\ref{Fig1}).
The Hamiltonian of the combined system can be written as \cite{Kheir1_2025,Kheir2_2025}
\begin{eqnarray}\label{H1}
\hat{H}(t) &=& \underbrace{\hbar \omega_0 \,(\hat{a}_1^\dag \hat{a}_1+\hat{a}_2^\dag \hat{a}_2)+\hbar k\,(\hat{a}_1\hat{a}_2^\dag+\hat{a}^\dag_1\hat{a}_2)}_{\hat{H}_S:\,\text{Main system}}\nn\\ 
&& + \underbrace{\hbar \omega_0 \,(\hat{b}^{\dagger} \hat{b}+\hat{c}^{\dagger} \hat{c})+\hbar k\,(\hat{b}\hat{c}^\dag+\hat{b}^\dag \hat{c})}_{\hat{H}_B:\,\text{Bath system}}\nn\\
&& + \underbrace{\hbar g(t)\,(\hat{a}_1\hat{b}^\dag+\hat{a}_1^\dag\hat{b}+\hat{a}_2\hat{c}^\dag+\hat{a}_2^\dag\hat{c})}_{\hat{H}_{SB} (t):\,\text{System-Bath interaction}}\nn\\
&& + \underbrace{\hbar\,f(t)\,(e^{-i\omega_L t}\hat{a}_1^\dag+e^{i\omega_L t}\hat{a}_1)}_{\text{Driving force}}.
\end{eqnarray}
where $g(t)$, is a time-dependent coupling function to be determined in the following. In a recent work \cite{Kheir2026}, it was shown that by engineering the system-bath coupling function as $\cos(G(t))\rightarrow e^{-\gamma t/2}$, where $\gamma$ is a dissipation coefficient and $G(t)=\int_0^t dt'\,g(t')$, to ensure unidirectional energy flow, the resulting reduced density matrices corresponding to prototype bosonic and fermionic systems exactly satisfy the Lindblad master equation. Also, in the subsection (\ref{initcoh}), we will show that the reduced density matrix corresponding to the main system, in zero temperature and in the absence of a driving force, satisfies the Lindblad master equation. The interested reader is referred to \cite{Kheir1_2025,Kheir2_2025} for further features of the present scheme.

By making use of the Bogoliubov transformations
\begin{eqnarray}\label{BTrs1}
  \hat{a}_1 &=& \frac{\hat{A}_1+\hat{A}_2}{\sqrt{2}},\,\,\,\, \hat{a}_2 = \frac{\hat{A}_1-\hat{A}_2}{\sqrt{2}},\nn \\
  \hat{b} &=& \frac{\hat{B}_1+\hat{B}_2}{\sqrt{2}},\,\,\,\, \hat{c} = \frac{\hat{B}_1-\hat{B}_2}{\sqrt{2}}.
\end{eqnarray}
The Hamiltonian Eq.~(\ref{H1}) can be rewritten as
\begin{eqnarray}\label{H2}
      \hat{H}(t) &=& \hbar (\omega_0+k)\,\hat{A}^{\dagger}_1 \hat{A}_1+\hbar\,g(t)\,[\hat{A}^\dag_1\,\hat{B}_1+\hat{A}_1\,\hat{B}^\dag_1]\nn\\
      && +\hbar(\omega_0+k)\hat{B}^\dag_1\hat{B}_1 +\frac{\hbar f(t)}{\sqrt{2}}\,(e^{-i\omega_L t}\,\hat{A}^\dag_1+e^{i\omega_L t}\,\hat{A}_1)\nn\\
      && + \hbar (\omega_0-k)\,\hat{A}^{\dagger}_2 \hat{A}_2+\hbar\,g(t)\,[\hat{A}^\dag_2\,\hat{B}_2+\hat{A}_2\,\hat{B}^\dag_2]\nn\\
      && +\hbar(\omega_0-k)\hat{B}^\dag_2\hat{B}_2
             +\frac{\hbar f(t)}{\sqrt{2}}\,(e^{-i\omega_L t}\,\hat{A}^\dag_2+e^{i\omega_L t}\,\hat{A}_2),
\end{eqnarray}
which can be decoupled by using another set of Bogoliubov transformation
\begin{eqnarray}\label{BTrs2}
\hat{A}_1=&\frac{\hat{C}_1+\hat{D}_1}{\sqrt{2}},\,\,\,\,\hat{B}_1=\frac{\hat{C}_1-\hat{D}_1}{\sqrt{2}},\nn\\
\hat{A}_2=&\frac{\hat{C}_2+\hat{D}_2}{\sqrt{2}},\,\,\,\,\hat{B}_2=\frac{\hat{C}_2-\hat{D}_2}{\sqrt{2}},
\end{eqnarray}
leading to a separation of the total Hamiltonian
\begin{eqnarray}\label{decoupled}
\hat{H}(t) && =\hbar(\omega_0+k+g(t))\,\hat{C}^\dag_1\hat{C}_1+\frac{\hbar f(t)}{2}\big[e^{-i\omega_L t}\,\hat{C}^\dag_1+e^{i\omega_L t}\,\hat{C}_1\big]\nn\\
        && \,\,+\hbar(\omega_0+k-g(t))\,\hat{D}^\dag_1\hat{D}_1+\frac{\hbar f(t)}{2}\big[e^{-i\omega_L t}\,\hat{D}^\dag_1+e^{i\omega_L t}\,\hat{D}_1\big]\nn\\
        && \,\,+\hbar(\omega_0-k+g(t))\,\hat{C}^\dag_2\hat{C}_2+\frac{\hbar f(t)}{2}\big[e^{-i\omega_L t}\,\hat{C}^\dag_2+e^{i\omega_L t}\,\hat{C}_2\big]\nn\\
        && \,\,+\hbar(\omega_0-k-g(t))\,\hat{D}^\dag_2\hat{D}_2+\frac{\hbar f(t)}{2}\big[e^{-i\omega_L t}\,\hat{D}^\dag_2+e^{i\omega_L t}\,\hat{D}_2\big].
\end{eqnarray}
Now from Heisenberg equations of motion one easily obtains
\begin{eqnarray}\label{4p}
\hat{C}_1(t)=& e^{-i\Omega_{1}^{+} (t)}\, \hat{C}_1(0)+\xi_1^{+} (t),\nn\\
\hat{C}_2(t)=& e^{-i\Omega_{2}^{+} (t)}\, \hat{C}_2(0)+\xi_2^{+} (t),\nn\\
\hat{D}_1(t)=& e^{-i\Omega_{1}^{-} (t)}\, \hat{D}_1(0)+\xi_{1}^{-} (t),\nn\\
\hat{D}_2(t)=& e^{-i\Omega_{2}^{-} (t)}\, \hat{D}_2(0)+\xi_{2}^{-} (t),
\end{eqnarray}
where for notational simplicity, we have defined
\begin{eqnarray} \label{defomega}
  \Omega^{\pm}_1 (t) =& (\omega_0+k) t\pm G(t),\nn\\
  \Omega^{\pm}_2 (t) =& (\omega_0-k) t\pm G(t),
\end{eqnarray}
\begin{eqnarray}\label{defkesi}
  \xi^{\pm}_1 (t) =& -\frac{i}{2}\int_0^t dt'\,e^{-i[\Omega^{\pm}_1 (t)-\Omega^{\pm}_1 (t')+\omega_L t']}\,f(t'),\nn\\
  \xi^{\pm}_2 (t) =& -\frac{i}{2}\int_0^t dt'\,e^{-i[\Omega^{\pm}_2 (t)-\Omega^{\pm}_2 (t')+\omega_L t']}\,f(t'),
\end{eqnarray}
\begin{equation}\label{Gt}
  G(t) =\int_0^t dt'\,g(t').
\end{equation}
Therefore, by applying inverse Bogoliubov transformations, we finally obtain the annihilation operators for the first and second primary oscillators in the Heisenberg picture
\begin{eqnarray}\label{finala1}
\hat{a}_1 (t) &=& p(t)\,\hat{a}_1 (0)+u(t)\,\hat{a}_2 (0)+q(t)\hat{b} (0)+h(t)\,\hat{c}(0)+ f_1 (t),\nn\\
\hat{a}_2 (t) &=& p(t)\,\hat{a}_2 (0)+u(t)\,\hat{a}_1 (0)+q(t)\,\hat{c} (0)+h(t)\hat{b}(0)+ f_2 (t),
\end{eqnarray}
where
\begin{eqnarray}\label{f1eq}
&&  f_1 (t)=\frac{1}{2}\big(\xi^{+}_1+\xi^{+}_2+\xi^{-}_1+\xi^{-}_2\big),\nn \\
&&\,\,\,\,\,\,\,\,\,\,\,\,\,\,= -i e^{-i\omega_0 t}\,\int_0^t dt'\,e^{i(\omega_0-\omega_L) t'}\cos[k(t-t')]\cos[G(t)-G(t')]\,f(t'),\\
&&  f_2 (t)=\frac{1}{2}\big(\xi^{+}_1-\xi^{+}_2+\xi^{-}_1-\xi^{-}_2\big),\nn\\
&&\,\,\,\,\,\,\,\,\,\,\,\,\,\,= - e^{-i\omega_0 t}\,\int_0^t dt'\,e^{i(\omega_0-\omega_L) t'}\sin[k(t-t')]\cos[G(t)-G(t')]\,f(t'),
\end{eqnarray}
and
\begin{eqnarray}
     p(t) &=& e^{-i\omega_0 t}\cos(G(t))\,\cos(kt),\nn\\
     u(t) &=& -i e^{-i\omega_0 t}\cos(G(t))\,\sin(kt),\nn\\
     q(t) &=& -i e^{-i\omega_0 t}\sin(G(t))\,\cos(kt),\nn\\
     h(t) &=& - e^{-i\omega_0 t}\sin(G(t))\,\sin(kt).
\end{eqnarray}
By setting $\cos^2 (G(t))=e^{-\gamma t}$ \cite{Kheir1_2025,Kheir2_2025,Kheir2026}, and taking trace over the degrees of freedom of the bath system, we find the average energy of the main oscillators in units of $\hbar\omega_0$ as
\begin{align} \label{e1t}
    \frac{E_1 (t)}{\hbar\omega_0} &=\la\hat{a}^\dag_1 (t)\hat{a}_1 (t)\ra \nn\\
    &= e^{-\gamma t}\,\big[\cos^2(k t)\,\la\hat{a}^\dag_1 (0)\hat{a}_1 (0)\ra+\sin^2(k t)\,\la\hat{a}^\dag_2(0)\hat{a}_2(0)\ra\big]\nn\\
    & \,\,\,\,\, +\bar{n}\,(1-e^{-\gamma t})+|f_1 (t)|^2,
\end{align}
\begin{align} \label{e2t}
    \frac{E_2 (t)}{\hbar\omega_0} &=\la\hat{a}^\dag_2 (t)\hat{a}_2 (t)\ra \nn\\
    &= e^{-\gamma t}\,\big[\cos^2(k t)\,\la\hat{a}^\dag_2 (0)\hat{a}_2 (0)\ra+\sin^2(k t)\,\la\hat{a}^\dag_1(0)\hat{a}_1(0)\ra\big]\nn\\
   & \,\,\,\,\, +\bar{n}\,(1-e^{-\gamma t})+|f_2 (t)|^2.
\end{align}
In Fig.~(\ref{Fig4}), the scaled energy of the right oscillator ($E_2 (t)/\hbar\omega_0$) is plotted as a function of the dimensionless parameter $kt$ for different values of the scaled detuning $\frac{\triangle}{k}=\frac{\omega_0-\omega_L}{k}$, at zero temperature. Also, the oscillators are assumed to be initially prepared in their ground states, and the energy is rescaled by the unit $E_0=F^2/k^2$, where $F$ is the constant amplitude of the external source ($f(t)=F$). At resonance, where $\triangle=\omega_0-\omega_L=0$, the energy oscillates within a restricted domain. However, for the value $\frac{\triangle}{k}=1$, the energy increases without bound.
\section{Husimi Q-functions}\label{Sec3}
The Husimi $Q$-function \cite{Knight2005} corresponding to a density matrix $\hat{\rho}$ is a positive definite distribution on phase space, defined by
\begin{equation}\label{Qfunction}
 Q(\alpha,\bar{\alpha})=\frac{1}{\pi}\la \alpha|\hat{\rho}|\alpha\ra,
\end{equation}
where $|\alpha\ra$ is a coherent state. In our case, the main system consists of the interacting oscillators described by the reduced density matrix $\hat{\rho}_S (t)$, and their corresponding coherent states are the tensor products $|\alpha_1,\alpha_2\ra=|\alpha_1\ra\otimes |\alpha_2\ra$, where $\hat{a}_i |\alpha_i\ra=\alpha_i |\alpha_i\ra,\,\,i=1,2$. The Husimi Q-function of the oscillators can then be obtained as (App. \ref{A})
\begin{eqnarray}\label{Husimi}
&& Q(\alpha_1,\bar{\alpha}_1,\alpha_2,\bar{\alpha}_2,t)=\frac{1}{\pi^2}\la\alpha_1,\alpha_2|\hat{\rho}^S (t)|\alpha_1,\alpha_2\ra\nn\\
&&= \frac{1}{\pi^2} \sum_{s_1,s_2=0}^\infty \Bigg\{\frac{(-1)^{s_1+s_2}}{s_1! s_2!}\,\tr\,\bigg[(\hat{a}^\dag_1 (t)-\bar{\alpha}_1)^{s_1}(\hat{a}_1 (t)-\alpha_1)^{s_1}
           (\hat{a}^\dag_2 (t)-\bar{\alpha}_2)^{s_2}(\hat{a}_2 (t)-\alpha_2)^{s_2}\,\hat{\rho}(0)\bigg]\Bigg\},\nn\\
\end{eqnarray}
where $\hat{\rho}(0)$ is the initial state of the total system. Straightforward calculations lead to (App. \ref{B})
\begin{eqnarray}\label{Husimi2}
  && Q(\alpha_1,\bar{\alpha}_1,\alpha_2,\bar{\alpha}_2,t)= \frac{1}{\pi^2} \sum_{s_1,s_2=0}^\infty \frac{(-1)^{s_1+s_2}}{s_1! s_2!}\partial^{s_1}_{\lambda_1}\partial^{s_1}_{\bar{\lambda}_1}\partial^{s_2}_{\lambda_2}\partial^{s_2}_{\bar{\lambda}_2}\Big\{e^{\zeta(\lambda_1,\lambda_2,\bar{\lambda}_1,\bar{\lambda}_2,t)}\nn\\
  && \times \,e^{\bar{n}\sin^2[G(t)](\lambda_1\bar{\lambda}_1+\lambda_2\bar{\lambda}_2)}\,\chi_1 (\lambda_1,\lambda_2,\bar{\lambda}_1,\bar{\lambda}_2)\,\chi_2 (\lambda_1,\lambda_2,\bar{\lambda}_1,\bar{\lambda}_2)\Big\}_{\tiny{\lambda_1,\lambda_2,\bar{\lambda}_1,\bar{\lambda}_2=0}},\nn\\
\end{eqnarray}
where for notational simplicity, we have defined 
\begin{eqnarray}\label{zeta}
  \zeta(\lambda_1,\lambda_2,\bar{\lambda}_1,\bar{\lambda}_2,t)&=&\lambda_1 (\bar{f}_1 (t)-\bar{\alpha}_1)+\bar{\lambda}_1 (f_1 (t)-\alpha_1)\nn\\
                                          && +\lambda_2 (\bar{f}_2 (t)-\bar{\alpha}_2)+\bar{\lambda}_2 (f_2 (t)-\alpha_2),
\end{eqnarray}
and the characteristic functions \cite{Knight2005} are defined by
\begin{eqnarray}\label{characteristics}
  && \chi_1 (\lambda_1,\lambda_2,\bar{\lambda}_1,\bar{\lambda}_2)=\tr_{a_1}\Big[e^{(\lambda_1\bar{p}+\lambda_2\bar{u})\hat{a}^\dag_1}e^{(\bar{\lambda}_1 p+\bar{\lambda}_2 u)\hat{a}_1}\,\hat{\rho}_1 (0)\Big],\nn \\
  && \chi_2 (\lambda_1,\lambda_2,\bar{\lambda}_1,\bar{\lambda}_2)=\tr_{a_2}\Big[e^{(\lambda_1\bar{u}+\lambda_2\bar{p})\hat{a}^\dag_2}e^{(\bar{\lambda}_1 u+\bar{\lambda}_2 p)\hat{a}_2}\,\hat{\rho}_2 (0)\Big].
\end{eqnarray}
\subsection{The oscillators are initially prepared in coherent states}\label{initcoh}
If the oscillators are initially prepared in the coherent state $\hat{\rho}_S (0)=|\alpha_0\ra_1\la \alpha_0|\otimes |\beta_0\ra_2\la \beta_0|$, then the Husimi Q-function of the oscillators can be easily found as
\begin{equation}\label{Qab}
  Q(\alpha_1,\bar{\alpha}_1,\alpha_2,\bar{\alpha}_2,t)=\frac{1}{\pi^2}\,\frac{1}{(1+\bar{n}\sin^2(G(t)))^2}\,\exp{\Big(-\frac{|\eta_1 (t)|^2+|\eta_2 (t)|^2}{1+\bar{n}\sin^2(G(t))}\Big)},
\end{equation}
where for notational simplicity we have defined
\begin{eqnarray}
  \eta_1 (t) &=& f_1 (t)+p (t)\alpha_0+u (t)\beta_0-\alpha_1,\nn \\
  \eta_2 (t) &=& f_2 (t)+u (t)\alpha_0+p (t)\beta_0-\alpha_2.
\end{eqnarray}
The locations of the maxima of the Husimi distribution function are defined by
\begin{eqnarray}\label{locationMax}
  \nu_1 (t) &=& f_1 (t)+p (t)\alpha_0+u (t)\beta_0,\nn \\
  \nu_2 (t) &=& f_2 (t)+u (t)\alpha_0+p (t)\beta_0.
\end{eqnarray}
In an environment at zero temperature, we set $\bar{n}=0$, and by making use of Eq.~(\ref{Qab}), we deduce that the evolved state is a separable coherent state given by
\begin{equation}\label{EvolvedCS}
  \hat{\rho}_S(t)=|\nu_1 (t) \ra\la \nu_1 (t) |\otimes |\nu_2 (t) \ra\la \nu_2 (t)|.
\end{equation}
Also, in the absence of a driving force ($f_1 (t)=f_2 (t)=0$), one can show that the density matrix Eq.~(\ref{EvolvedCS}) fulfills the Lindblad master equation (App. \ref{C})
\begin{eqnarray}\label{Lindblad1}
 \dot{\hat{\rho}}_S(t)=-\frac{i}{\hbar}[\hat{H}_S (t),\hat{\rho}_S (t)] &+\gamma\Big(\hat{a}_1 \hat{\rho}_S (t) \hat{a}^\dag_1-\frac{1}{2}\big\{\hat{a}^\dag_1 \hat{a}_1,\hat{\rho}_S (t)\big\}\Big)\nn\\
 &+\gamma\Big(\hat{a}_2 \hat{\rho}_S (t) \hat{a}^\dag_2-\frac{1}{2}\big\{\hat{a}^\dag_2 \hat{a}_2,\hat{\rho}_S (t)\big\}\Big).
\end{eqnarray}
\subsubsection{Reduced density matrices}\label{reduced}
\begin{figure}[h] 
    \centering
    \includegraphics[width=0.3\textwidth]{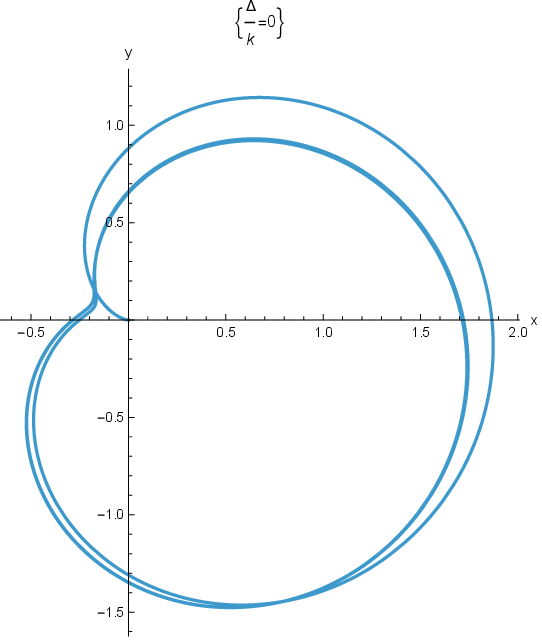}
    \includegraphics[width=0.3\textwidth]{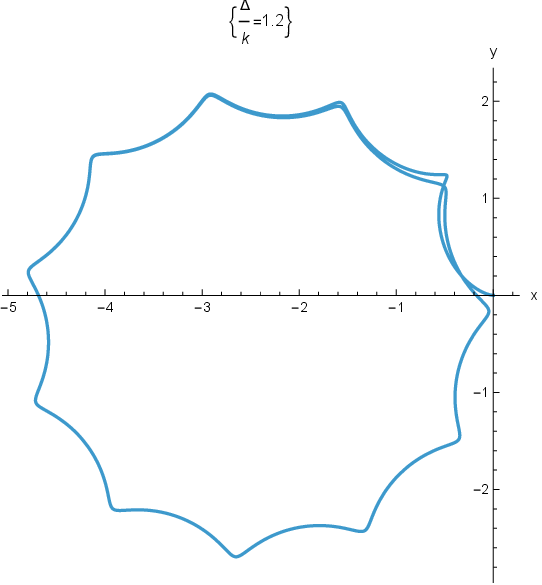}
    \includegraphics[width=0.6\textwidth]{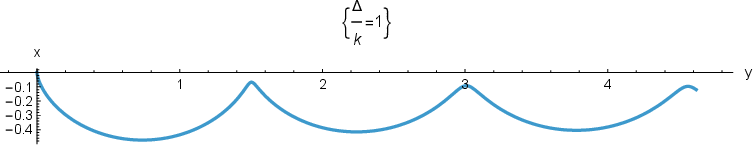}
    \caption{The locations of the maximum of the Husimi function of the right oscillator at zero temperature and in the presence of an external source with constant amplitude $F$ and frequency $\omega_L$ for the scaled detuning values $\frac{\triangle}{k}=0,\,1.2,\,1$, and $\frac{F}{k}=1$. For $\frac{\triangle}{k}=1$, the path is not closed and the energy of the oscillator, which is proportional to the squared distance from the origin, increases without bound.}
    \label{Fig3}
\end{figure}

From Eq.~(\ref{Qab}) we deduce that the Husimi distribution is separable in phase space and can be written as
\begin{equation}\label{Q1Q2}
  Q(\alpha_1,\bar{\alpha}_1,\alpha_2,\bar{\alpha}_2,t)=Q(\alpha_1,\bar{\alpha}_1,t)\,Q(\alpha_2,\bar{\alpha}_2,t),
\end{equation}
where
\begin{equation}\label{Q12}
  Q(\alpha_i,\bar{\alpha}_i,t)=\frac{1}{\pi}\,\frac{1}{1+\bar{n}\sin^2(G(t))}\,\exp{\Big(-\frac{|\eta_i (t)|^2}{1+\bar{n}\sin^2(G(t))}\Big)},\,\,\,i=1,2,
\end{equation}
is the Husimi function of the ith oscillator ($i=1,2$). It should be noted that the Husimi functions depend on the initial conditions of both oscillators. At zero temperature, we have $\bar{n}=0$, and
\begin{eqnarray}
Q(\alpha_i,\bar{\alpha}_i,t) &=& \frac{1}{\pi}e^{-|\eta_i (t)|^2}= \frac{1}{\pi}e^{-|\nu_i (t)-\alpha_i|^2},\nn\\
&=& \frac{1}{\pi}\la \alpha_i|\nu_i (t)\ra\la \nu_i (t)|\alpha_i\ra,
\end{eqnarray}
where $\nu_i$ are given by Eqs.~(\ref{locationMax}). Therefore, the final state of the oscillators is a separable coherent state given by $|\nu_1 (t)\ra\la \nu_1 (t)|\otimes |\nu_2 (t)\ra\la \nu_2 (t)|$. In Fig.~(\ref{Fig3}), the locations of the maxima of the Husimi functions for the right oscillator are depicted at zero temperature in the presence of an external force with constant amplitude $F$ and frequency $\omega_L$, for both resonance ($\frac{\triangle}{k}=0$) and off-resonance $(\frac{\triangle}{k}\neq 0$) cases. For $\frac{\triangle}{k}=1$, the path is not closed and the energy of the oscillator, which is proportional to the squared distance from the origin, increases without bound.

Let us concentrate on the second oscillator described by the Husimi function $Q(\alpha_2,\bar{\alpha}_2,t)$. For notational simplicity, from now on we write $\alpha$ instead of $\alpha_2$. To find the reduced density matrix corresponding to the second oscillator, we make use of the following formula \cite{Knight2005}
\begin{equation}\label{Elegant}
  \rho_{mn} (t)=\frac{\pi}{\sqrt{m!\,n!}}\,\Big[\frac{\partial^m}{\partial{\bar{\alpha}}^m}\frac{\partial^n}{\partial{\alpha}^n}
  \Big(Q(\alpha,\bar{\alpha},t)\,e^{\bar{\alpha}\alpha}\Big)\Big]_{\alpha=\bar{\alpha}=0}.
\end{equation}

Thus, the Husimi function can be considered as a generating function for the elements of the density matrix in the number state basis $\{|n\ra\}_{n=0}^\infty$. To find the components of the reduced density matrix corresponding to the second oscillator in the basis of number states, we substitute the reduced Husimi function Eq.~(\ref{Q12}) into Eq.~(\ref{Elegant}) and obtain (App. \ref{D})
\begin{eqnarray}\label{romn}
  \rho_{mn} (t) &=& \sqrt{\frac{n!}{m!}}\,\frac{e^{-\frac{|r_t|^2}{\sigma_t}}(\sigma_t-1)^m (\bar{r}_t)^{n-m}}{(\sigma_t)^{n+1}}\sum_{k=0}^m\,\binom{m}{k}\frac{1}{(n-m+k)!}\,\Big(\frac{|r_t|^2}{\sigma_t-1}\Big)^k,\nn\\
\end{eqnarray}
where
\begin{equation}\label{rt}
r_t=f_2 (t)+u (t)\alpha_0+p (t)\beta_0,
\end{equation}
\begin{equation}
\sigma_t=1+\bar{n}\,\sin^2(G(t)).
\end{equation}\label{sigmat}
The population probabilities can be obtained from the diagonal elements of the density matrix in Eq.~(\ref{romn}), i.e., $P_n (t)=\rho_{nn} (t)$, given by
\begin{eqnarray}
&& P_n (t)=\exp{\Big(-\frac{|r_t|^2}{\bar{n}\sin^2(G(t))+1}\Big)}\,\frac{(\bar{n}\sin^2(G(t)))^n}{(\bar{n}\sin^2(G(t))+1)^{n+1}}\,L_n \Big(-\frac{|r_t|^2}{\bar{n}\sin^2(G(t))}\Big),\nn\\
\end{eqnarray}
where $L_n (\cdot)$ is the Laguerre polynomial of degree $n$. At zero temperature, $\bar{n}=0$, and $\sigma_t\rightarrow 1$, in this case, Eq.~(\ref{romn}) becomes
\begin{eqnarray}\label{romn00}
  \rho^{T=0}_{mn} (t) &=& \sqrt{\frac{1}{m! n!}}\,r^m_t\,\bar{r}^n_t\,e^{-|r_t|^2},\nn\\
                      &=& \la m|r_t\ra\la r_t|n\ra,
\end{eqnarray}
therefore, $\hat{\rho}^{T=0}(t)=|r_t\ra\la r_t|$, where $|r_t\ra$ is a coherent state with the parameter $r_t$ given by Eq.~(\ref{rt}).
\subsection{The oscillators are initially prepared in number states}\label{initnum}
If the oscillators are initially prepared in number states $\hat{\rho}_S (0)=|N\ra_1\la N|\otimes |0\ra_2\la 0|$, then (App. \ref{E})
\begin{equation}\label{QN}
Q_N (\alpha_1,\bar{\alpha}_1,\alpha_2,\bar{\alpha}_2,t)=\frac{1}{\pi^2\sigma^2_t} e^{-\frac{1}{\sigma_t}\big[|f_1-\alpha_1|^2+|f_2-\alpha_2|^2\big]}\,\big[e(t)\big]^N\,L_N\bigg(-\frac{|h(t)|^2}{e(t)}\bigg),
\end{equation}
where, for simplicity, we have defined
\begin{eqnarray}\label{et}
  e(t) &=& 1-\frac{|p (t)|^2+|u (t)|^2}{\sigma_t}=1-\frac{\cos^2(G(t))}{\sigma_t},\nn\\
  h(t) &=& \frac{1}{\sigma_t}\bigg((\alpha_1-f_1 (t))\bar{p} (t)+(\alpha_2-f_2 (t))\bar{u} (t)\bigg).
\end{eqnarray}
\subsubsection{Example: N=1}
For the case $N=1$, the initial density matrix of the two oscillators are given by
\begin{equation}\label{ExampleN1}
\hat{\rho}_S (0)=|1\ra_1\la 1|\otimes |0\ra_2\la 0|,
\end{equation}
At zero temperature and in the absence of an external driving force, we set $f_1 (t)=f_2 (t)=\bar{n}=0$ in Eq.~(\ref{QN}). Using Eq.~(\ref{Husimi}), the time-evolved density matrix is obtained as (App. \ref{F})
\begin{eqnarray}\label{N1density}
  \hat{\rho}_S (t) &=& (1-e^{-\gamma t})\,|0\ra_1\la 0|\otimes |0\ra_2\la 0|\nn\\
  && + e^{-\gamma t}\cos^2(k t)\,|1\ra_1\la 1|\otimes |0\ra_2\la 0|\nn\\
  && + e^{-\gamma t}\sin^2(k t)\,|0\ra_1\la 0|\otimes |1\ra_2\la 1|\nn\\
  && + \frac{i}{2}\,e^{-\gamma t}\sin(2k t)\big(|1\ra_1\la 0|\otimes |0\ra_2\la 1|-|0\ra_1\la 1|\otimes |1\ra_2\la 0|\big).
\end{eqnarray}
Thus, at time $t$, the system remains in its initial state with probability $e^{-\gamma t}\cos^2(k t)$, transfers the excitation to the second oscillator with probability $e^{-\gamma t}\sin^2(k t)$, and decays to the joint ground state with probability $1-e^{-\gamma t}$. The last term in Eq.~(\ref{N1density}) represent coherency which decays due to the interaction with the medium. In the long-time limit, thermalization is achieved, and the system approaches the final state $|0\ra_1\la 0|\otimes |0\ra_2\la 0|$.
\subsection{Generating function for $Q_N (\alpha_1,\bar{\alpha}_1,\alpha_2,\bar{\alpha}_2,t)$}
\begin{figure}[h]
    \centering
    \includegraphics[width=0.4\textwidth]{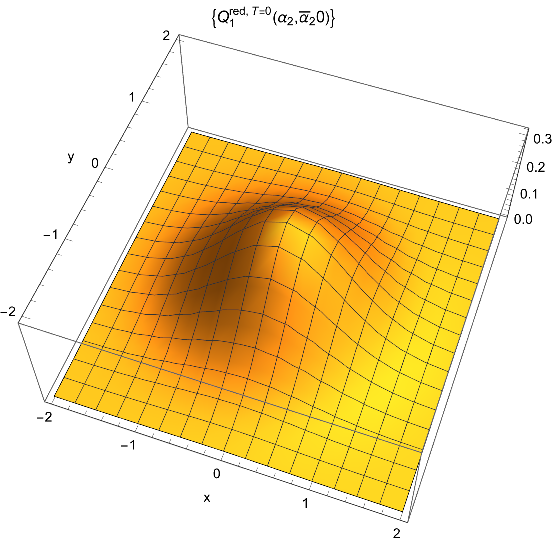}
    \includegraphics[width=0.4\textwidth]{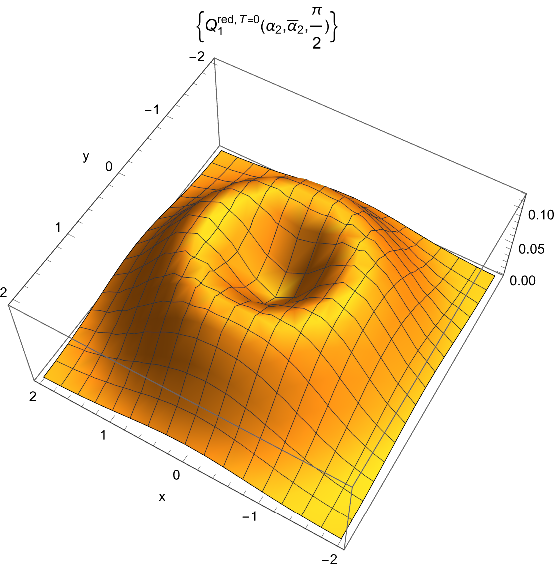}
        \caption{(a) Husimi function $Q^{red}_1 (\alpha_2,\bar{\alpha}_2,0)$ at zero temperature. (b) Husimi function $Q^{red}_1 (\alpha_2,\bar{\alpha}_2,kt=\frac{\pi}{2})$ at zero temperature. Over time, after several oscillations at the peak, the Husimi function eventually settles into the left state (a).}
    \label{Fig4}
\end{figure}

A generating function for $Q_N (\alpha_1,\bar{\alpha}_1,\alpha_2,\bar{\alpha}_2,t)$ can be found by multiplying each side of Eq.~(\ref{QN}) by $y^N$, summing over $N$, and using the identity
\begin{equation}\label{ylaguerre}
\sum_{N=0}^\infty y^N\,L_N (x)=\frac{e^{-\frac{xy}{1-y}}}{1-y},
\end{equation}
which yields
\begin{eqnarray}\label{GenQ}
&&  Q_y(\alpha_1,\bar{\alpha}_1,\alpha_2,\bar{\alpha}_2,t) = \sum_{N=0}^\infty y^N\,Q_N (\alpha_1,\bar{\alpha}_1,\alpha_2,\bar{\alpha}_2,t),\nn \\
&&  = \frac{1}{\pi^2\sigma^2_t} e^{-\frac{1}{\sigma_t}\big[|f_1-\alpha_1|^2+|f_2-\alpha_2|^2\big]}\, \frac{1}{1-y\,e(t)}\,e^{\frac{y\,|w(t)|^2}{1-y\,e(t)}}.
\end{eqnarray}
Consequently,
\begin{equation}\label{GenQ2}
  Q_N (\alpha_1,\bar{\alpha}_1,\alpha_2,\bar{\alpha}_2,t)=\frac{1}{N!}\frac{\partial^N}{\partial y^N}  Q_y(\alpha_1,\bar{\alpha}_1,\alpha_2,\bar{\alpha}_2,t)\Big|_{y=0}.
\end{equation}
The reduced Husimi function corresponding to the right oscillator can be obtained by integrating over $\alpha_1$
\begin{eqnarray}\label{reducedHu}
  Q^{red}_N (\alpha_2,\bar{\alpha}_2,t) &=& \int d^2\alpha_1\,Q_N (\alpha_1,\bar{\alpha}_1,\alpha_2,\bar{\alpha}_2,t),\nn\\
  &=& \frac{1}{N!}\frac{\partial^N}{\partial y^N} Q^{red}_y (\alpha_2,\bar{\alpha}_2,t)\Big|_{y=0},
\end{eqnarray}
where we have defined
\begin{equation}\label{Qyred}
  Q^{red}_y (\alpha_2,\bar{\alpha}_2,t)=\int d^2\alpha_1\,Q_y(\alpha_1,\bar{\alpha}_1,\alpha_2,\bar{\alpha}_2,t).
\end{equation}
In Eq.~(\ref{Qyred}), $Q_y (\cdots)$ is quadratic in its variables $\alpha_i,\,\bar{\alpha}_i$, therefore, integration over $\alpha_1$ can be achieved straightforwardly, leading to (App. \ref{G})
\begin{equation}\label{Qyfinal}
   Q^{red}_y (\alpha_2,\bar{\alpha}_2,t)=\frac{1}{\pi}\,e^{-\frac{1}{\sigma_t}|\alpha_2-f_2(t)|^2}\frac{1}{A(y)}
   e^{\frac{|u(t)|^2|\alpha_2-f_2(t)|^2}{\sigma^2_t}B(y)},
\end{equation}
where we have defined
\begin{eqnarray}
  A(y) &=& \sigma_t-y[e(t)\sigma_t+|p (t)|^2],\nn \\
  B(y) &=& \frac{|p (t)|^2y^2}{\sigma_t^2(1-e(t)y)^2}+\frac{y}{1-e(t)y}.
\end{eqnarray}
The reduced Husimi function $Q^{red}_N (\alpha_2,\bar{\alpha}_2,t)$ can then be obtained using Eq.~(\ref{reducedHu}). As an example, consider $N=1$; that is, the initial state of the oscillators is $\hat{\rho}_S (0)=|1\ra_1\la 1|\otimes|0\ra_2\la 0|$. From Eq.~(\ref{reducedHu}), we find
\begin{eqnarray}\label{example}
 && Q^{red}_1 (\alpha_2,\bar{\alpha}_2,t) = \frac{1}{1!}\frac{\partial^N}{\partial y^N}\, Q^{red}_y (\alpha_2,\bar{\alpha}_2,t)\Big|_{y=0},\nn\\
 &&  =\frac{1}{\pi} e^{-\frac{1}{\sigma_t}|\alpha_2-f_2 (t)|^2}\Big(\frac{|u (t)|^2|\alpha_2-f_2 (t)|^2}{\sigma^3_t}+\frac{[e(t)\sigma_t+|p (t)|^2]}{\sigma^2_t}\Big).
\end{eqnarray}
In Fig.~(\ref{Fig4}), the Husimi function $Q^{red}_1 (\alpha_2,\bar{\alpha}_2,t)$ is plotted in the absence of an external source ($f(t)=0$) and at zero temperature ($\bar{n}=0$), for $kt=0$ and $kt=\pi/2$.
In the long-time regime, the Husimi function in Eq.~(\ref{QN}) becomes a separable thermal state
\begin{equation}\label{QNsep}
 Q(\alpha_1,\bar{\alpha}_1,\alpha_2,\bar{\alpha}_2,t)=\frac{1}{\pi (1+\bar{n})} e^{-|f_1 (t)-\alpha_1|^2}\,\frac{1}{\pi (1+\bar{n})} e^{-|f_2 (t)-\alpha_2|^2},
\end{equation}
as expected.
\section{Generalization}\label{Sec4}
\begin{figure}[h]
\centering
\includegraphics[scale=0.15]{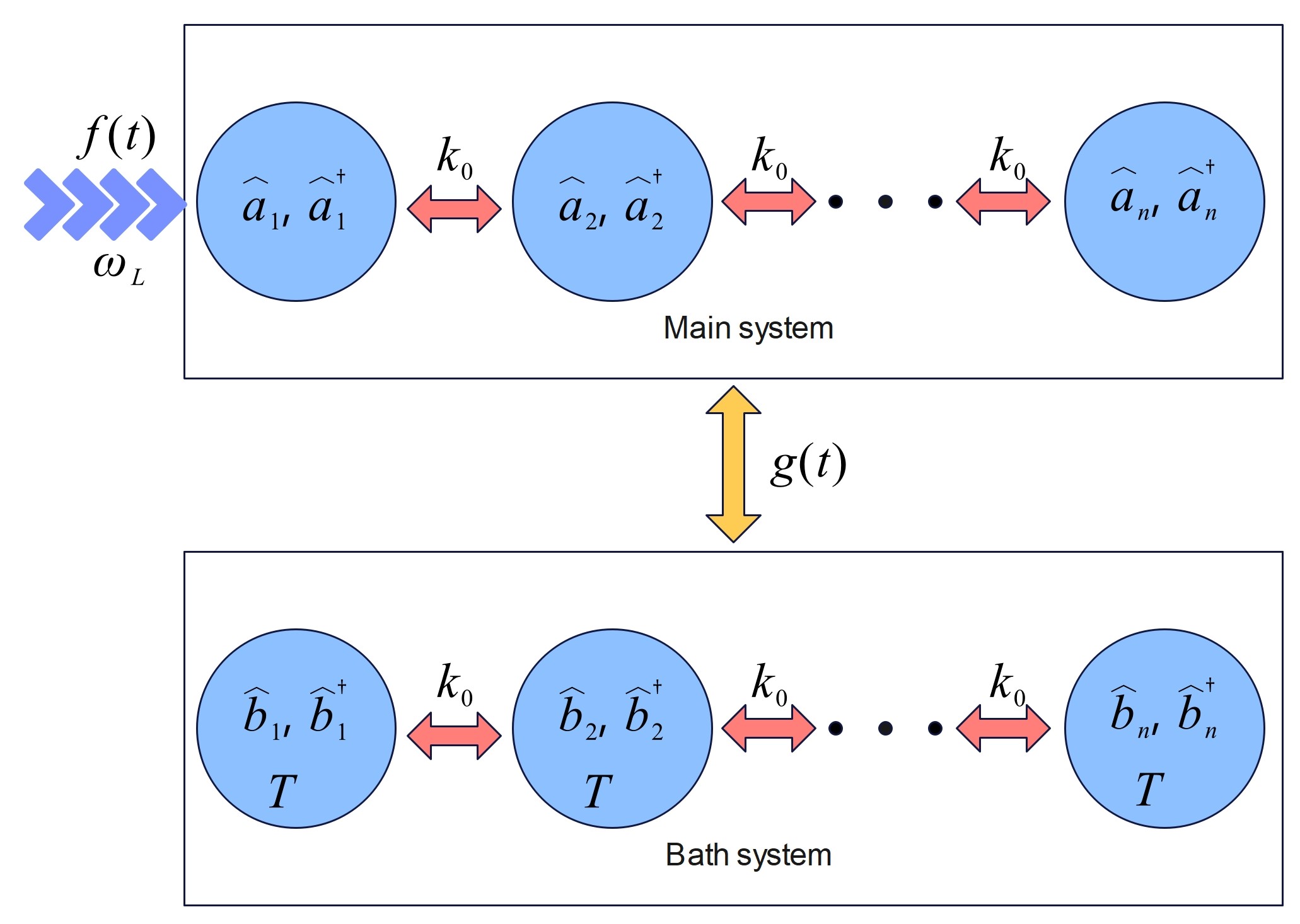}
\caption{(Color online) $n$-interacting oscillators in a medium driven by a classical field $f(t)$ applied to the first oscillator of the main system. The time-dependent coupling functions $g(t)$ couple the main system to the bath oscillators. The strength of the interaction between the oscillators in the main system and the bath is determined by the parameter $k_0$.}\label{Fig5}
\end{figure}

\begin{figure}
    \centering
    \includegraphics[width=0.4\textwidth]{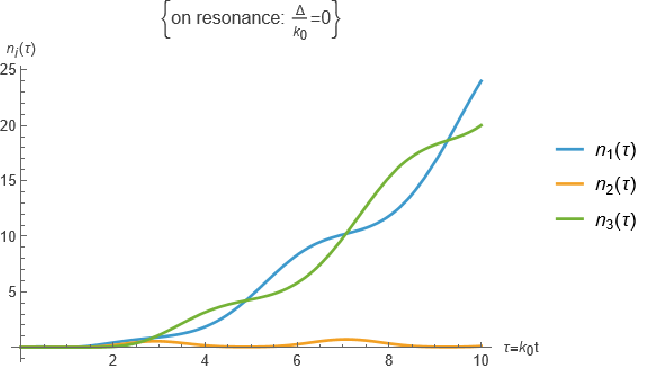}
    \includegraphics[width=0.4\textwidth]{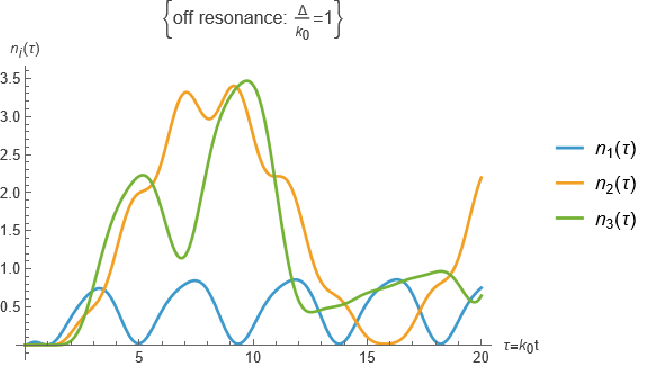}
        \caption{(a) Scaled excitations $n_i (\tau)$ as a function of the dimensionless parameter $\tau=k_0 t$ for $n=3$ at resonance ($\frac{\Delta}{k_0}=0$). (b) Scaled excitations $n_i (\tau)$ as a function of the dimensionless parameter $\tau=k_0 t$ for $n=3$ off resonance ($\frac{\Delta}{k_0}=1$).}
    \label{Fig6}
\end{figure}

In this section, we generalize the system to $n$-interacting oscillators in a medium driven by a classical field, see Fig.~(\ref{Fig5}).
The Hamiltonian of the combined system can be written as
\begin{eqnarray}\label{HforN}
  \hat{H} &=& \hbar\omega_0\sum_{k=1}^n \big(\hat{a}^\dag_k \hat{a}_k+\hat{b}^\dag_k \hat{b}_k\big)+\hbar k_0 \sum_{k=1}^{n-1} \big(\hat{a}_k \hat{a}^\dag_k+\hat{a}^\dag_k \hat{a}_k\big)\nn\\
   &+& \hbar g(t)\,\sum_{k=1}^n \big(\hat{a}_k \hat{b}^\dag_k+\hat{a}^\dag_k \hat{b}_k\big)+\hbar f(t)\big(e^{-i\omega_L t}\,\hat{a}^\dag_1+e^{i\omega_L t}\,\hat{a}_1\big).
\end{eqnarray}
The Hamiltonian in Eq.~(\ref{HforN}) can be rewritten in matrix form as
\begin{eqnarray}\label{HforNMatrix}
&& \hat{H}=\left[
                \begin{array}{cccc}
                  \hat{a}^\dag_1 & \hat{a}^\dag_2 & \cdots & \hat{a}^\dag_n \\
                \end{array}
              \right]\mathcal{M}\left[
                                  \begin{array}{c}
                                    \hat{a}_1 \\
                                    \hat{a}_2 \\
                                    \vdots \\
                                    \hat{a}_n \\
                                  \end{array}
                                \right] + \left[
                \begin{array}{cccc}
                  \hat{b}^\dag_1 & \hat{b}^\dag_2 & \cdots & \hat{b}^\dag_n \\
                \end{array}
              \right]\mathcal{M}\left[
                                  \begin{array}{c}
                                    \hat{b}_1 \\
                                    \hat{b}_2 \\
                                    \vdots \\
                                    \hat{b}_n \\
                                  \end{array}
                                \right]\nn\\
&& +\hbar g(t)\left(
                \begin{array}{cccc}
                  \hat{a}^\dag_1 & \hat{a}^\dag_2 & \cdots & \hat{a}^\dag_n \\
                \end{array}
              \right)\left[
                                  \begin{array}{c}
                                    \hat{b}_1 \\
                                    \hat{b}_2 \\
                                    \vdots \\
                                    \hat{b}_n \\
                                  \end{array}
                                \right] + \hbar g(t)\left[
                \begin{array}{cccc}
                  \hat{b}^\dag_1 & \hat{b}^\dag_2 & \cdots & \hat{b}^\dag_n \\
                \end{array}
              \right]\left[
                                  \begin{array}{c}
                                    \hat{a}_1 \\
                                    \hat{a}_2 \\
                                    \vdots \\
                                    \hat{a}_n \\
                                  \end{array}
                                \right]\nn\\
&& +\hbar f(t)\big(e^{-i\omega_L t}\,\hat{a}^\dag_1+e^{i\omega_L t}\,\hat{a}_1\big),\nn\\
\end{eqnarray}
where $\mathcal{M}$ is a three-diagonal matrix given by
\begin{equation}\label{3diagonalM}
  \mathcal{M}=\left[
                \begin{array}{cccccc}
                  \hbar\omega_0 & \hbar k_0 & \cdots & 0 & \cdots & 0 \\
                  \hbar k_0 & \hbar\omega_0 & \hbar k_0 & 0 & \cdots & 0 \\
                  0 & \hbar k_0 & \hbar\omega_0 & \hbar k_0 & \cdots & \vdots \\
                  0 & 0 & \hbar k_0 & \ddots & \ddots & 0 \\
                  \vdots & \ddots & \ddots & \ddots & \ddots & \hbar k_0 \\
                  0 & 0 & \cdots & 0 & \hbar k_0 &  \hbar\omega_0 \\
                \end{array}
              \right].
\end{equation}
The three-diagonal matrix $\mathcal{M}$ can be diagonalized using its eigenvalues and eigenvectors \cite{Noschese2013}. Let us denote the orthogonal transformation matrix that diagonalizes $\mathcal{M}$ by $T$, then $T^t \mathcal{M} T=diag(\hbar\lambda_1,\cdots,\hbar\lambda_n)$ where $\{\hbar \lambda_k\}_{k=1}^n$ are eigenvalues of $\mathcal{M}$ given by
\begin{equation}\label{eigenvaluesM}
  \hbar\lambda_k=\hbar\omega_0+2\hbar k_0\cos\Big(\frac{k\pi}{n+1}\Big),\,\,\,\,k=1,2,\cdots,n,
\end{equation}
with the corresponding eigenvectors
\begin{equation}\label{eigenvectorM}
  |\hbar\lambda_k\ra=\sqrt{\frac{2}{n+1}}\left(
                                           \begin{array}{c}
                                             \sin\big(\frac{k\pi}{n+1}\big) \\
                                             \sin\big(\frac{2k\pi}{n+1}\big) \\
                                             \vdots \\
                                             \sin\big(\frac{nk\pi}{n+1}\big) \\
                                           \end{array}
                                         \right).
\end{equation}
Therefore, the explicit form of the transformation matrix $T$ is
\begin{equation}\label{transformT}
T=\left(
  \begin{array}{cccc}
    \sqrt{\frac{2}{n+1}}\sin\big(\frac{\pi}{n+1}\big) & \sqrt{\frac{2}{n+1}}\sin\big(\frac{2\pi}{n+1}\big) & \cdots & \sqrt{\frac{2}{n+1}}\sin\big(\frac{n\pi}{n+1}\big) \\
    \sqrt{\frac{2}{n+1}}\sin\big(\frac{2\pi}{n+1}\big) & \sqrt{\frac{2}{n+1}}\sin\big(\frac{4\pi}{n+1}\big) & \cdots & \sqrt{\frac{2}{n+1}}\sin\big(\frac{2n\pi}{n+1}\big) \\
    \vdots & \vdots & \cdots & \vdots \\
    \sqrt{\frac{2}{n+1}}\sin\big(\frac{n\pi}{n+1}\big) & \sqrt{\frac{2}{n+1}}\sin\big(\frac{2n\pi}{n+1}\big) & \cdots & \sqrt{\frac{2}{n+1}}\sin\big(\frac{n^2\pi}{n+1}\big) \\
  \end{array}
\right),
\end{equation}
or equivalently
\begin{equation}\label{compoT}
  T_{ij}=\sqrt{\frac{2}{n+1}}sin\Big(\frac{ij\pi}{n+1}\Big).
\end{equation}
Now we define new bosonic operators
\begin{eqnarray}\label{newops}
&&  \left[
     \begin{array}{c}
       \hat{A}_1 \\
       \vdots \\
       \hat{A}_n \\
     \end{array}
   \right]
   = T^t \left[
             \begin{array}{c}
               \hat{a}_1 \\
               \vdots \\
               \hat{a}_n \\
             \end{array}
           \right],
    \nn\\
&&  \left[
     \begin{array}{c}
       \hat{B}_1 \\
       \vdots \\
       \hat{B}_n \\
     \end{array}
   \right]
   = T^t \left[
             \begin{array}{c}
               \hat{b}_1 \\
               \vdots \\
               \hat{b}_n \\
             \end{array}
           \right],
\end{eqnarray}
satisfying the commutation relations
\begin{eqnarray}\label{commutators}
&&  [\hat{A}_i, \hat{A}^\dag_j]=[\hat{B}_i , \hat{B}^\dag_j]=\delta_{ij},\nn \\
&&  [\hat{A}_i,\hat{A}_j] = [\hat{B}_i,\hat{B}_j]=0,\nn\\
&&  [\hat{A}_i,\hat{B}_j]=[\hat{A}_i,\hat{B}^\dag_j]=0.
\end{eqnarray}
The Haniltonian in Eq.~(\ref{HforNMatrix}), expressed in terms of the new operators, becomes
\begin{eqnarray}\label{HAB}
\hat{H} &=& \sum_{k=1}^n \Big[\hbar\lambda_k \big(\hat{A}^\dag_k\hat{A}_k+\hat{B}^\dag_k\hat{B}_k\big)+\hbar g(t)\big(\hat{A}^\dag_k\hat{B}_k+\hat{A}_k\hat{B}^\dag_k\big)\nn\\
        && \,\,\,\,\,\,\,+ \hbar f(t)\,T_{k1}\big(e^{-i\omega_L t}\,\hat{A}^\dag_k+e^{i\omega_L t}\,\hat{A}_k\big)\Big],\nn\\
        &=& \sum_{k=1}^n \hat{H}_k,
\end{eqnarray}
where each $\hat{H}_k$ has the same structure as Eq.~(\ref{H2}), given by
\begin{eqnarray}\label{Hk2}
  \hat{H}_k &=& \hbar\lambda_k\,(\hat{A}^\dag \hat{A}_k+\hat{B}^\dag \hat{B}_k)+\hbar g(t)\,(\hat{A}^\dag_k \hat{B}_k+\hat{A}_k \hat{B}^\dag_k)\nn \\
   && + \hbar f(t)\,T_{k1}(e^{-i \omega_L t}\,\hat{A}^\dag_k+e^{i\omega_L t}\,\hat{A}_k).
\end{eqnarray}
By making use of the Bogoliubov transformations
\begin{eqnarray}\label{genBogo}
  \hat{A}_k &=& \frac{\hat{C}_k+\hat{D}_k}{\sqrt{2}},\nn \\
  \hat{B}_k &=& \frac{\hat{C}_k-\hat{D}_k}{\sqrt{2}},
\end{eqnarray}
The Hamiltonian $\hat{H}_k$ can be rewritten as
\begin{eqnarray}\label{Hkcd}
  \hat{H}_k &=& \hbar(\lambda_k+g(t))\,\hat{C}^\dag_k\hat{C}_k+\frac{\hbar f(t)\,T_{k1}}{\sqrt{2}}\,\Big(e^{-i\omega_L t}\,\hat{C}^\dag_k+e^{i\omega_L t}\,\hat{C}_k\Big)\nn\\
  &+& \hbar(\lambda_k-g(t))\,\hat{D}^\dag_k\hat{D}_k+\frac{\hbar f(t)\,T_{k1}}{\sqrt{2}}\,\Big(e^{-i\omega_L t}\,\hat{D}^\dag_k+e^{i\omega_L t}\,\hat{D}_k\Big).\nn\\
\end{eqnarray}
Now, by solving Heisenberg equations, one readily finds
\begin{eqnarray}\label{CandD}
  \hat{C}_k (t) &=& e^{-i\eta^+_k (t)}\,\hat{C}_k (0)+\frac{\hbar T_{k1}}{\sqrt{2}}\int_0^t e^{-i[\eta^{+}_k (t)-\eta^{+}_k (t')]}\,f(t') e^{-i\omega_L t'} dt',\nn \\
 \hat{D}_k (t) &=& e^{-i\eta^{-}_k (t)}\,\hat{D}_k (0)+\frac{\hbar T_{k1}}{\sqrt{2}}\int_0^t e^{-i[\eta^{-}_k (t)-\eta^{-}_k (t')]}\,f(t') e^{-i\omega_L t'} dt',\nn\\
\end{eqnarray}
where we have defined $\eta^{\pm}_k (t)=\lambda_k t\pm G(t)$. Using Eqs.~(\ref{genBogo}), we obtain
\begin{eqnarray}\label{Ak}
  \hat{A}_k (t) &=& \frac{1}{2} \Big(\big[e^{-i\eta^{+}_k (t)}+e^{-i\eta^{-}_k (t)}\big]\,\hat{A}_k (0)+\big[e^{-i\eta^{+}_k (t)}-e^{-i\eta^{-}_k (t)}\big]\,\hat{B}_k (0)\Big)\nn\\
  && +\frac{\hbar}{2}\,T_{k1}\int_0^t \big(e^{-i[\eta^{+}_k (t)-\eta^{+}_k (t')]}+e^{-i[\eta^{-}_k (t)-\eta^{-}_k (t')]}\big) e^{-i\omega_L t'}f(t') dt'.\nn\\
\end{eqnarray}
Finally, using the inverse transforms
\begin{eqnarray}\label{newops}
&&  \left[
     \begin{array}{c}
       \hat{a}_1 \\
       \vdots \\
       \hat{a}_n \\
     \end{array}
   \right]
   = T \left[
             \begin{array}{c}
               \hat{A}_1 \\
               \vdots \\
               \hat{A}_n \\
             \end{array}
           \right],
    \nn\\
&&  \left[
     \begin{array}{c}
       \hat{b}_1 \\
       \vdots \\
       \hat{b}_n \\
     \end{array}
   \right]
   = T \left[
             \begin{array}{c}
               \hat{B}_1 \\
               \vdots \\
               \hat{B}_n \\
             \end{array}
           \right],
\end{eqnarray}
we obtain
\begin{eqnarray}\label{smalla}
  \hat{a}_l (t)&=&\sum_{j,k}^n \Big\{\big(T_{lk}u_k (t) T^t_{kj}\big)\,\hat{a}_j (0)+\big(T_{lk}v_k (t) T^t_{kj}\big)\,\hat{b}_j (0)\Big\} + \sum_{k=1}^n T_{lk} \zeta_k(t),
\end{eqnarray}
where for convenience, we have defined
\begin{eqnarray}
  u_k (t) &=& \frac{1}{2} \big(e^{-i\eta^{+}_k (t)}+e^{-i\eta^{-}_k (t)}\big),\nn \\
  v_k (t) &=& \frac{1}{2} \big(e^{-i\eta^{+}_k (t)}-e^{-i\eta^{-}_k (t)}\big),
\end{eqnarray}
\begin{eqnarray}\label{zetak}
  \zeta_k(t) &=& \frac{\hbar}{2}\int_0^t\, dt' \big(T_{k1}e^{-i[\eta^{+}_k (t)-\eta^{+}_k (t')]}+T_{1k}e^{-i[\eta^{-}_k (t)-\eta^{-}_k (t')]}\big) e^{-i\omega_L t'}f(t'),\nn\\
  &=& \hbar\,T_{k1} \int_0^t\,dt'\, e^{-i\lambda_k (t-t')}e^{-i\omega_L t'}f(t')\cos[G(t)-G(t')].
\end{eqnarray}
The Eq.~(\ref{smalla}) can be rewritten in compact matrix form as
\begin{equation}\label{compacta}
 \hat{a}(t)=\tilde{u}(t)\,\hat{a}(0)+\tilde{v}(t)\,\hat{b}(0)+\tilde{\zeta}(t),
\end{equation}
where
\begin{eqnarray}\label{Matrices}
&&  \tilde{u}(t) = T\,\left(
                        \begin{array}{ccc}
                          u_1 (t) & \cdots & 0 \\
                          \vdots & \ddots & \vdots \\
                          0 & \cdots & u_n (t) \\
                        \end{array}
                      \right)\,T^t,\nn
   \\
&&  \tilde{v}(t) = T\,\left(
                        \begin{array}{ccc}
                          v_1 (t) & \cdots & 0 \\
                          \vdots & \ddots & \vdots \\
                          0 & \cdots & v_n (t) \\
                        \end{array}
                      \right)\,T^t,\nn
   \\
&&  \left(
     \begin{array}{c}
       \tilde{\zeta}_1 (t) \\
       \vdots \\
       \tilde{\zeta}_n (t) \\
     \end{array}
   \right)
   = T \left(
           \begin{array}{c}
              \zeta_1 (t) \\
             \vdots \\
              \zeta_n (t) \\
           \end{array}
         \right)\nn\\
&& \hat{a} (t)=\left(
               \begin{array}{c}
                 \hat{a}_1 (t) \\
                 \vdots \\
                 \hat{a}_n (t) \\
               \end{array}
             \right).
\end{eqnarray}
As an example, let the oscillators of the system be initially prepared in their ground states and the bath oscillators be held at zero temperature
\begin{equation}\label{initN}
  \hat{\rho}(0)=\underbrace{|0\ra_1\la 0|\otimes |0\ra_2\la 0|\otimes\cdots \otimes |0\ra_n\la 0|}_{\text{System}}\otimes \underbrace{|0\ra_B\la 0|}_{\text{Bath}},
\end{equation}
then
\begin{equation}\label{adogaN}
  \la \hat{a}^\dag_i (t)\hat{a}_i (t)\ra=|\psi_i(t)|^2,
\end{equation}
where
\begin{eqnarray}\label{zetafunction}
 && \psi_i(t) =\frac{\hbar\,e^{-i\omega_0 t}}{k_0}\,\int_0^t dt'\,f(t')e^{i\Delta t'}\cos[G(t)-G(t')]\,\phi_i (t-t'),\nn\\
 && \phi_i (t-t')=\sum_{k=1}^n e^{-2ik_0(t-t')\cos[\frac{k\pi}{n+1}]}\,T_{ik}T_{k1}.\nn\\
\end{eqnarray}
and $\Delta=\omega_0-\omega_L$ is detuning. In Fig.~(\ref{Fig6}), for three oscillators ($n=3$), the scaled excitations
\begin{equation}\label{excitations}
  n_i (\tau)=\frac{k_0^4}{\hbar^2 F^2}\la \hat{a}^\dag_i (\tau)\hat{a}_i (\tau)\ra,
\end{equation}
are plotted as a function of the dimensionless parameter $\tau=k_0 t$ for $n=3$ in the presence of an external source with frequency $\omega_L$ and constant amplitude $F_0$ for on-resonance $(\frac{\Delta}{k_0}=0)$ and off-resonance $(\frac{\Delta}{k_0}=1)$ cases.
\section{Conclusions}\label{conclusions}
In this work, we have developed a comprehensive analytical framework for investigating the nonequilibrium quantum dynamics of interacting harmonic oscillators coupled to a modelled thermal reservoir and driven by external classical fields. The core of our approach lies in the exact diagonalization of the total system-reservoir Hamiltonian via successive Bogoliubov transformations. For the case of two coupled oscillators with an external source applied to the first, we obtained closed-form expressions for the energies of the individual oscillators. The phase-space characterization provided further insight into the quantum state of the system. Explicit Husimi $Q$-functions for both oscillators, along with their reduced forms, were derived for initial states prepared as separable coherent and number states. The loci of the Q-function maxima, illustrated for various parameter values at zero temperature, offer a clear geometric visualization of the state evolution. Additionally, we obtained explicit expressions for the reduced density matrix elements in the number-state basis. A central result is the demonstration that, in the absence of external driving and at zero temperature, the reduced dynamics exactly reduces to the Lindblad master equation-thereby establishing a direct and rigorous link between our microscopic diagonalization scheme and the phenomenological Markovian description commonly used in quantum optics. We also introduced a generating function for the $Q$-function, providing a compact tool for computing correlation functions and moments of arbitrary order. The formalism was further generalized to a system of $n$ interacting oscillators embedded in a thermal medium and driven by an external classical field. Analytical expressions for the energy of each oscillator were derived, and the specific case $n=3$ was examined in detail, demonstrating the scalability and versatility of our method for more complex quantum networks and many-body systems.
\newpage

\appendix

\section{Derivation of Eqs.~(\ref{Husimi})}\label{A}
\begin{eqnarray}
   && Q(\alpha_1,\bar{\alpha}_1,\alpha_2,\bar{\alpha}_2,t) = \frac{1}{\pi^2}\la\alpha_1,\alpha_2|\hat{\rho}^S (t)|\alpha_1,\alpha_2\ra,\nn \\
   &&= \frac{1}{\pi^2}\la\alpha_1,\alpha_2|\tr_B [\hat{\rho} (t)]|\alpha_1,\alpha_2\ra,\nn \\
   &&= \frac{1}{\pi^2}\,\tr_S\,\tr_B\,\big[|\alpha_1,\alpha_2\ra\la\alpha_1,\alpha_2|\,\hat{\rho} (t)\big],\nn \\
   &&= \frac{1}{\pi^2}\,\tr\,\big[|\alpha_1,\alpha_2\ra\la\alpha_1,\alpha_2|\,\hat{U}(t)\hat{\rho}(0)\hat{U}^\dag (t)\big],\nn \\
   &&= \frac{1}{\pi^2}\,\tr\,\big[\hat{U}^\dag (t)|\alpha_1,\alpha_2\ra\la\alpha_1,\alpha_2|\hat{U}(t)\,\hat{\rho}(0)\Big],\nn \\
   &&= \frac{1}{\pi^2}\,\tr\,\big[\hat{U}^\dag (t) \hat{D}_1(\alpha_1)|0\ra_{a_1}\la 0|\hat{D}^\dag_1 (\alpha_1)\otimes\hat{D}_2 (\alpha_2)|0\ra_{a_2}\la 0|\hat{D}^\dag_2 (\alpha_2)\hat{U}(t)\hat{\rho} (0)\Big],\nn \\
   &&= \frac{1}{\pi^2}\,\sum_{s_1,s_2=0}^\infty\frac{(-1)^{s_1+s_2}}{s_1! s_2!}\tr\,\Bigg\{\hat{U}^\dag (t) \hat{D}_1(\alpha_1)(\hat{a}^\dag_1)^{s_1} (\hat{a}_1)^{s_1}\hat{D}^\dag_1(\alpha_1)\nn\\
   && \,\,\,\,\,\,\,\,\,\,\,\,\,\,\hspace{4cm}\hat{D}_2(\alpha_2)(\hat{a}^\dag_2)^{s_2} (\hat{a}_2)^{s_2}\hat{D}^\dag_2(\alpha_2)\hat{U}(t)\hat{\rho} (0)\Bigg\},\nn\\
   &&= \frac{1}{\pi^2}\,\sum_{s_1,s_2=0}^\infty\frac{(-1)^{s_1+s_2}}{s_1! s_2!}\tr\,\bigg\{(\hat{a}_1^\dag (t)-\bar{\alpha}_1)^{s_1}
   (\hat{a}_1 (t)-\alpha_1)^{s_1}\nn\\
   &&\hspace{5cm}(\hat{a}_2^\dag (t)-\bar{\alpha}_2)^{s_2} (\hat{a}_2 (t)-\alpha_2)^{s_2}\hat{\rho}(0)\Bigg\},\nn\\
\end{eqnarray}
where, for notational simplicity, we have defined the displacement operators
\begin{eqnarray}
  \hat{D}_1(\alpha_1) &=& e^{\alpha_1\hat{a}_1^\dag-\bar{\alpha}_1\hat{a}_1},\nn \\
  \hat{D}_2(\alpha_2) &=& e^{\alpha_2\hat{a}_2^\dag-\bar{\alpha}_2\hat{a}_2},\nn
\end{eqnarray}
and made use of the following relations
\begin{eqnarray}
  \hat{D}_1(\alpha_1)\hat{a}_1 \hat{D}_1^\dag(\alpha_1)&=& \hat{a}_1-\alpha_1,\nn \\
  \hat{D}_1(\alpha_1)\hat{a}_1^\dag \hat{D}_1^\dag(\alpha_1)&=& \hat{a}_1^\dag-\bar{\alpha}_1,\nn \\
  \hat{D}_2(\alpha_2)\hat{a}_2 \hat{D}_2^\dag(\alpha_2)&=& \hat{a}_2-\alpha_2,\nn \\
  \hat{D}_2(\alpha_2)\hat{a}_2^\dag \hat{D}_1^\dag(\alpha_2)&=& \hat{a}_2^\dag-\bar{\alpha}_2,\nn.
\end{eqnarray}
\section{Derivation of Eq.~(\ref{Husimi2})}\label{B}
To obtain Eq.~(\ref{Husimi2}), we rewrite Eq.~(\ref{Husimi}) as
\begin{eqnarray}
 && Q(\alpha_1,\hat{\alpha}_1,\alpha_2,\hat{\alpha}_2,t)=\frac{1}{\pi^2}\,\sum_{s_1=0,s_2=0}^\infty\frac{(-1)^{s_1+s_2}}{s_1! s_2!}\partial^{s_1}_{\lambda_1}\partial^{s_1}_{\bar{\lambda}_1}\partial^{s_2}_{\lambda_2}\partial^{s_2}_{\bar{\lambda}_2}\nn\\
 && \tr\,\Big[e^{\lambda_1 (\hat{a}^\dag_1 (t)-\bar{\alpha}_1)}e^{\bar{\lambda}_1 (\hat{a}_1 (t)-\alpha_1)}e^{\lambda_2 (\hat{a}^\dag_2 (t)-\bar{\alpha}_2)}e^{\bar{\lambda}_2 (\hat{a}_2 (t)-\alpha_2)}\hat{\rho}(0)\Big]_{\lambda_1,\lambda_2,\bar{\lambda}_1,\bar{\lambda}_2=0}\nn\\
 && =\frac{1}{\pi^2}\,\sum_{s_1=0,s_2=0}^\infty\frac{(-1)^{s_1+s_2}}{s_1! s_2!}\partial^{s_1}_{\lambda_1}\partial^{s_1}_{\bar{\lambda}_1}\partial^{s_2}_{\lambda_2}\partial^{s_2}_{\bar{\lambda}_2}e^{-\lambda_1 \bar{\alpha}_1-\bar{\lambda}_1\alpha_1\lambda_2\bar{\alpha}_2-\bar{\lambda}_2 \alpha_2}\nn\\
 && \tr\,\Big\{e^{\lambda_1[\bar{p} (t)\hat{a}^\dag_1+\bar{u} (t)\hat{a}^\dag_2+\bar{q} (t)\hat{b}^\dag+\bar{h} (t)\hat{c}^\dag +\bar{f}_1 (t)]}e^{\bar{\lambda}_1 [p (t)\hat{a}_1+u(t)\hat{a}_2+q(t)\hat{b}+h(t)\hat{c}+f_1 (t)]}\nn\\
 && e^{\lambda_2[\bar{p} (t)\hat{a}^\dag_2+\bar{u} (t)\hat{a}^\dag_1+\bar{q}(t)\hat{c}^\dag+\bar{h} (t)\hat{b}^\dag +\bar{f}_2 (t)]}e^{\bar{\lambda}_2 [p (t)\hat{a}_2+ u(t)\hat{a}_1+q(t)\hat{c}+h(t)\hat{b}+f_2 (t)]} \Big\}_{\tiny{\lambda_1,\lambda_2,\bar{\lambda}_1,\bar{\lambda}_2=0}},\nn\\
 && =\frac{1}{\pi^2}\,\sum_{s_1=0,s_2=0}^\infty\frac{(-1)^{s_1+s_2}}{s_1! s_2!}\partial^{s_1}_{\lambda_1}\partial^{s_1}_{\bar{\lambda}_1}\partial^{s_2}_{\lambda_2}\partial^{s_2}_{\bar{\lambda}_2}\Bigg\{e^{\lambda_1 (\bar{f}_1-\bar{\alpha}_1)+\bar{\lambda}_1 (f_1-\alpha_1)
 +\lambda_2 (\bar{f}_2-\bar{\alpha}_2)+\bar{\lambda}_2 (f_2-\alpha_2)}\nn\\
 && \tr_{a_1}\Big(e^{\lambda_1 \bar{p}\hat{a}^\dag_1}e^{\bar{\lambda}_1 p\hat{a}_1}e^{\lambda_2 \bar{u}\hat{a}^\dag_1}e^{\bar{\lambda}_2 u\hat{a}_1}\hat{\rho}_1 (0)\Big)\tr_{a_2}\Big(e^{\lambda_1 \bar{u}\hat{a}^\dag_2} e^{\bar{\lambda}_1 u\hat{a}_2}e^{\lambda_2 \bar{p}\hat{a}^\dag_2}e^{\bar{\lambda}_2 p\hat{a}_2}\hat{\rho}_2 (0)\Big)\nn\\
 && \tr_{b}\Big(e^{\lambda_1 \bar{q}\hat{b}^\dag}e^{\bar{\lambda}_1 q\hat{b}}e^{\lambda_2 \bar{h}\hat{b}^\dag}e^{\bar{\lambda}_2 h\hat{b}}\hat{\rho}_b (0)\Big)\tr_{c}\Big(e^{\lambda_1 \bar{w}\hat{c}^\dag}e^{\bar{\lambda}_1 h(t)\hat{c}}e^{\lambda_2 \bar{q}\hat{c}^\dag}e^{\bar{\lambda}_2 q\hat{c}}\hat{\rho}_c (0)\Big)\Bigg\},\nn\\
 && =\frac{1}{\pi^2}\,\sum_{s_1=0,s_2=0}^\infty\frac{(-1)^{s_1+s_2}}{s_1! s_2!}\partial^{s_1}_{\lambda_1}\partial^{s_1}_{\bar{\lambda}_1}\partial^{s_2}_{\lambda_2}\partial^{s_2}_{\bar{\lambda}_2}\Bigg\{
 e^{\lambda_1(\bar{f}_1-\bar{\alpha}_1)+\bar{\lambda}_1(f_1-\alpha_1)+\lambda_2 (\bar{f}_2-\bar{\alpha}_2)+\bar{\lambda}_2 (f_2-\alpha_2)}\nn\\
 &&\times\,e^{\bar{n}\sin^2[G(t)] (\lambda_1 \bar{\lambda}_1+\lambda_2 \bar{\lambda}_2)}\,\tr\Big(e^{(\lambda_1\bar{p}+\lambda_2\bar{u})\hat{a}^\dag_1}e^{(\bar{\lambda}_1 p+\bar{\lambda}_2 u)\hat{a}_1}\hat{\rho}_1 (0)\Big)\nn\\
 &&\times\,\tr\Big(e^{(\lambda_1\bar{u}+\lambda_2\bar{p})\hat{a}^\dag_2}e^{(\bar{\lambda}_1 u+\bar{\lambda}_2 p)\hat{a}_2}\hat{\rho}_2 (0)\Big)\Bigg\}_{\tiny{\lambda_1,\lambda_2,\bar{\lambda}_1,\bar{\lambda}_2=0}},\nn
\end{eqnarray}
where we made use of the following identities
\begin{eqnarray}
 && e^{x\hat{a}}\,e^{y\hat{a}^\dag} = e^{y\hat{a}^\dag}\, e^{x\hat{a}}\,e^{xy},\nn \\
 && \la n|e^{x\hat{a}^\dag} e^{y\hat{a}}|n\ra=L_n (-xy),\nn\\
 && \sum_{n=0}^\infty y^n\,L_{n}(x)=\frac{e^{\frac{-xy}{1-y}}}{1-y},\nn \\
 && q\bar{h}+\bar{q}h=0,\nn\\
 && u\bar{p}+\bar{u}p=0.\nn
\end{eqnarray}
\section{Derivation of Eq.~(\ref{Lindblad1})}\label{C}
In the absence of a driving force, we have
\begin{eqnarray}
  \nu_1 (t) &=& p(t)\alpha_0+u(t)\beta_0,\nn\\
  \nu_2 (t) &=& u(t)\alpha_0+p(t)\beta_0. 
\end{eqnarray}
If we set $\cos(G(t))=e^{-\frac{\gamma t}{2}}$, then 
\begin{eqnarray}
  p(t) &=& e^{-i\omega_0 t}\,e^{-\frac{\gamma t}{2}}\cos(k t),\nn\\
  u(t) &=& -i e^{-i\omega_0 t}\,e^{-\frac{\gamma t}{2}}\sin(k t),
\end{eqnarray}
with the time derivatives
\begin{eqnarray}
  \dot{p}(t) &=& -(i\omega_0+\frac{\gamma}{2})\,p(t)-i k\,u(t),\nn\\
  \dot{u}(t) &=& -(i\omega_0+\frac{\gamma}{2})\,u(t)-i k\,p(t),\nn\\
  \dot{\bar{p}}(t) &=& (i\omega_0-\frac{\gamma}{2})\,\bar{p}(t)+i k\,\bar{u}(t),\nn\\
  \dot{\bar{u}}(t) &=& (i\omega_0-\frac{\gamma}{2})\,\bar{u}(t)+i k\,\bar{p}(t).
\end{eqnarray}
Also, by taking the time derivative of the coherent state $|\nu_i\ra\,\,\,(i=1,2)$, we obtain
\begin{eqnarray}
  \frac{d}{d t}\,|\nu_i (t)\ra &=& \frac{d}{d t}\Big(e^{-\frac{1}{2}|\nu_i (t)|^2}\sum_{l=0}^\infty \frac{(\nu_i (t))^l}{\sqrt{l!}}\,|l\ra\Big),\nn \\
    &=& -\frac{1}{2}\frac{d}{d t}\big(|\nu_i (t)|^2\big)\,|\nu_i (t)\ra+\dot{\nu}_i (t)\,e^{-\frac{1}{2}|\nu_i (t)|^2}\sum_{l=0}^\infty \frac{l\,(\nu_i (t))^{l-1}}{\sqrt{l!}}\,|l\ra \nn\\
    &=& -\frac{1}{2}\frac{d}{d t}\big(|\nu_i (t)|^2\big)\,|\nu_i (t)\ra+\dot{\nu}_i (t)\,\hat{a}^\dag_i\,|\nu_i (t)\ra, 
\end{eqnarray}
and similarly
\begin{eqnarray}
  \frac{d}{d t}\,\la\nu_i (t)|= -\frac{1}{2}\frac{d}{d t}\big(|\nu_i (t)|^2\big)\,\la\nu_i (t)|+\dot{\bar{\nu}}_i (t)\,\la\nu_i (t)|\,\hat{a}_i.
\end{eqnarray}
Now by taking the time derivative of both sides of Eq.~(\ref{EvolvedCS}), and using the following identities
\begin{eqnarray}
&& |\nu_1 (t)|^2+|\nu_2 (t)|^2 = e^{-\gamma t}\,\big(|\alpha_0|^2+|\beta_0|^2\big),\nn \\
&&  \dot{\nu}_1 (t) = -(i\omega_0 +\frac{\gamma}{2})\,\nu_1 (t)-ik\,\nu_2 (t),\nn \\
&&  \dot{\nu}_2 (t) = -(i\omega_0 +\frac{\gamma}{2})\,\nu_2 (t)-ik\,\nu_1 (t),\nn \\
&&  \hat{a}_i\, |\nu_i (t)\ra = \nu_i (t)\, |\nu_i (t)\ra,\nn\\
&& \la \nu_i (t)|\,\hat{a}^\dag_i=\bar{\nu}_i (t)\,\la \nu_i (t)|, 
\end{eqnarray}
we finally obtain Eq.~(\ref{Lindblad1}).
\section{Derivation of Eq.~(\ref{romn})}\label{D}
We have
\begin{eqnarray}
  \rho_{mn} (t) &=& \frac{\pi}{\sqrt{m! n!}}\partial^m_{\bar{\alpha}}\partial^n_{\alpha}|\Bigg\{\frac{e^{\bar{\alpha}\alpha}}{\pi\sigma_t}e^{-\frac{|r_t-\alpha|^2}{\sigma_t}}
  \Bigg\}_{\bar{\alpha}=\alpha=0},\nn \\
   &=& \frac{1}{\sqrt{m! n!}\,\sigma_t}\partial^m_{\bar{\alpha}}\partial^n_{\alpha}\Bigg\{e^{(1-\frac{1}{\sigma_t})\bar{\alpha}\alpha-\frac{|r_t|^2}{\sigma_t}+
   \frac{1}{\sigma_t}(r_t\,\bar{\alpha}+\bar{r}_t\alpha)}\Bigg\}_{\bar{\alpha}=\alpha=0},\nn \\
   &=& \frac{e^{-\frac{|r_t|^2}{\sigma_t}}}{\sqrt{m! n!}\,\sigma_t}\partial^m_{\bar{\alpha}}\Bigg\{\Big(\big(1-\frac{1}{\sigma_t}\big)\bar{\alpha}+\frac{\bar{r}_t}{\sigma_t}\Big)^n
   \,e^{\frac{r_t}{\sigma_t}\,\bar{\alpha}}\Bigg\}_{\bar{\alpha}=0},\nn \\
   &=& \frac{e^{-\frac{|r_t|^2}{\sigma_t}}}{\sqrt{m! n!}\,\sigma_t}\sum_{k=0}^m\,\binom{m}{k}\,\partial^{m-k}_{\bar{\alpha}}\Bigg\{\Big(\big(1-\frac{1}{\sigma_t}\big)\bar{\alpha}+\frac{\bar{r_t}}{\sigma_t}\Big)^n
   \Bigg\}\,\partial^k_{\bar{\alpha}}\big[\,e^{\frac{r_t}{\sigma_t}\bar{\alpha}}\big]\Bigg|_{\bar{\alpha}=0},\nn \\
   &=& \sqrt{\frac{n!}{m!}}\,\frac{e^{-\frac{|r_t|^2}{\sigma_t}}(\sigma_t-1)^m (\bar{r}_t)^{n-m}}{(\sigma_t)^{n+1}}\sum_{k=0}^m\,\binom{m}{k}\frac{1}{(n-m+k)!}\,\Big(\frac{|r_t|^2}{\sigma_t-1}\Big)^k.\nn\\
\end{eqnarray}
\section{Derivation of Eq.~(\ref{QN})}\label{E}
We have
\begin{eqnarray}\label{D1}
\hat{\rho} (0) &=& \hat{\rho}_S (0)\otimes \hat{\rho}_B (0),\nn\\
\hat{\rho}_S (0) &=& |N\ra_1\la N|\otimes |0\ra_2\la 0|,\nn\\
   &=& \frac{1}{N!} \partial^N_{\gamma}\partial^N_{\bar{\gamma}}\,\Big\{e^{\bar{\gamma}\gamma}\,|\gamma\ra_1\la \gamma|\otimes |0\ra_2\la 0|\Big\}_{\gamma=\bar{\gamma}=0},
\end{eqnarray}
where $\hat{\rho}_B (0)$ is a thermal state. By inserting Eqs.~(\ref{D1}) into Eq.~(\ref{Husimi}), we have
\begin{eqnarray}
&& Q_N(\alpha_1,\bar{\alpha}_1,\alpha_2,\bar{\alpha}_2,t)=\frac{1}{N!} \partial^N_{\gamma}\partial^N_{\bar{\gamma}}\,\Bigg\{\frac{e^{\bar{\gamma}\gamma}}{\pi^2} \sum_{s_1,s_2=0}^\infty \Bigg\{\frac{(-1)^{s_1+s_2}}{s_1! s_2!}\nn\\
&&\times\,\tr_S\,\bigg[(\hat{a}^\dag_1 (t)-\bar{\alpha}_1)^{s_1}(\hat{a}_1 (t)-\alpha_1)^{s_1}
           (\hat{a}^\dag_2 (t)-\bar{\alpha}_2)^{s_2}(\hat{a}_2 (t)-\alpha_2)^{s_2}\nn\\
&& |\gamma\ra_1\la \gamma|\otimes |0\ra_2\la 0|\otimes \hat{\rho}_B (0)\bigg]\Bigg\}\Bigg\}_{\gamma=\bar{\gamma}=0},\nn\\
&& =\frac{1}{N!} \partial^N_{\gamma}\partial^N_{\bar{\gamma}}\,\Bigg\{\frac{e^{\bar{\gamma}\gamma}}{\pi^2 \sigma^2_t}\,e^{-\frac{1}{\sigma_t}\Big(|f_1 (t)+p(t)\gamma-\alpha_1|^2+|f_2 (t)+u(t)\gamma-\alpha_2|^2\Big)}\Bigg|_{\gamma=\bar{\gamma}=0},
\end{eqnarray}
where, in the last line, we made use of the results of the (\ref{initcoh}). Therefore,
\begin{eqnarray}
&& Q_N(\alpha_1,\bar{\alpha}_1,\alpha_2,\bar{\alpha}_2,t)=\frac{1}{N!\pi^2\sigma^2_t} e^{-\frac{1}{\sigma_t}\big[|f_1-\alpha_1|^2+|f_2-\alpha_2|^2\big]}\nn\\
&&\times\, \partial^N_{\gamma}\partial^N_{\bar{\gamma}}\Bigg\{e^{\big(1-\frac{|p(t)|^2+|u(t)|^2}{\sigma_t}\big)
\bar{\gamma}\gamma+h\bar{\gamma}+\bar{h}\gamma}\Bigg\}_{\gamma=\bar{\gamma}=0},\nn\\
&=& \frac{1}{N!\pi^2\sigma^2_t} e^{-\frac{1}{\sigma_t}\big[|f_1-\alpha_1|^2+|f_2-\alpha_2|^2\big]}\nn\\
&&\times\,\partial^N_\gamma\Bigg\{\bigg[\big(1-\frac{|p(t)|^2+|u(t)|^2}{\sigma_t}\big)\,
\gamma+h\bigg]^N\,e^{\bar{h}\gamma}\Bigg\}_{\gamma=0},\nn\\
&&= \frac{1}{N!\pi^2\sigma^2_t} e^{-\frac{1}{\sigma_t}\big[|f_1-\alpha_1|^2+|f_2-\alpha_2|^2\big]}\nn\\
&&\times\,\sum_{k=0}^N\binom{N}{k}\partial^{N-k}_\gamma\,\bigg(\bigg(1-\frac{|p(t)|^2+|u(t)|^2}{\sigma_t}\bigg)\,
\gamma+h\bigg)^N\,\partial^k_\gamma\big(e^{\bar{h}\gamma}\big)\Bigg|_{\gamma=0},\nn\\
&&= \frac{1}{N!\pi^2\sigma^2_t} e^{-\frac{1}{\sigma_t}\big[|f_1-\alpha_1|^2+|f_2-\alpha_2|^2\big]}\nn\\
&&\times\,\sum_{k=0}^N\binom{N}{k}\frac{|h|^{2k}N!}{k!}\,\Big(1-\frac{|p(t)|^2+|u(t)|^2}{\sigma_t}\Big)^{N-k},\nn\\
&&= \frac{1}{\pi^2\sigma^2_t} e^{-\frac{1}{\sigma_t}\big[|f_1-\alpha_1|^2+|f_2-\alpha_2|^2\big]}\,\big(e(t)\big)^N\,L_N\bigg(-\frac{|h|^2}{e(t)}\bigg),\nn\\
\end{eqnarray}
where we made use of the definition of the Laguerre polynomials
\begin{equation}\label{Lager}
  L_N (x)=\sum_{k=0}^N \binom{N}{k}\frac{(-x)^k}{k!}.
\end{equation}
\section{Derivation of Eq.~(\ref{N1density})}\label{F}
By using the definition in Eq.~(\ref{Husimi}), we have
\begin{eqnarray}
 \la \alpha_1, \alpha_2| \hat{\rho}_S (t) |\alpha_1,_2\ra &=& e^{-|\alpha_1|^2-|\alpha_2|^2}\,\big(1-e^{-\gamma t}+|\alpha_1|^2 |p(t)|^2+|\alpha_2|^2 |u(t)|^2+\alpha_1 \bar{\alpha}_2\,\bar{p}(t)u(t)+\bar{\alpha}_1\alpha_2\,p(t)\bar{u}(t)\big) \nn\\
 &=& \la \alpha_1, \alpha_2|\big((1-e^{-\gamma t})\,|0\ra_{12}\la 0|+e^{-\gamma t}\cos^2(k t)\,\hat{a}^\dag_1 |0\ra_{12}\la 0|\,\hat{a}_1+ e^{-\gamma t}\sin^2(k t)\,\hat{a}^\dag_2 |0\ra_{12}\la 0|\,\hat{a}_2\nn\\
 && - (\frac{i}{2}\,e^{-\gamma t}\sin(2k t)\,\hat{a}^\dag_2\,|0\ra_{12}\la 0|\,\hat{a}_1 + (\frac{i}{2}\,e^{-\gamma t}\sin(2k t)\,\hat{a}^\dag_1\,|0\ra_{12}\la 0|\,\hat{a}_2\big)|\alpha_1,\alpha_2\ra,
\end{eqnarray}
where for notational simplicity, we have defined $|0\ra_{12}\la 0|=|0\ra_1\la 0|\otimes |0\ra_2\la 0|$. Applying the standard coherent-state identities $\hat{a}_i |\alpha_i\ra=\alpha_i |\alpha_i\ra$, $\la \alpha_i|\hat{a}^\dag_i=\bar{\alpha}_i \la \alpha_i|$, we directly recover Eq.~(\ref{N1density}).

By making use of the following relations
\begin{eqnarray}
&& [\hat{a}^\dag_{1} \hat{a}_1 + \hat{a}^\dag_{2} \hat{a}_2, \hat{\rho}_S (t)]=0,\nn\\
&& \{\hat{a}^\dag_{1} \hat{a}_1 + \hat{a}^\dag_{2} \hat{a}_2, \hat{\rho}_S (t)\}=2e^{-\gamma t}\cos^2(k t)|1\ra_1\la 1|\otimes|0\ra_2\la 0|+2e^{-\gamma t}\sin^2(k t)|0\ra_1\la 0|\otimes|1\ra_2\la 1|\nn\\
&& \hspace{3.5cm}+i e^{-\gamma t}\sin(2k t)\big[|1\ra_1\la 0|\otimes|0\ra_2\la 1|-|0\ra_1\la 1|\otimes|1\ra_2\la 0|\big],\nn\\
&& [\hat{a}^\dag_{1} \hat{a}_2 + \hat{a}_{1} \hat{a}^\dag_2, \hat{\rho}_S (t)]=e^{-\gamma t}\cos(2k t)\big[|0\ra_1\la 1|\otimes|1\ra_2\la 0|-|1\ra_1\la 0|\otimes|0\ra_2\la 1|\big]\nn\\
&& \hspace{3.5cm} +ie^{-\gamma t}\sin(2k t)\big[|0\ra_1\la 0|\otimes|1\ra_2\la 1|-|1\ra_1\la 1|\otimes|0\ra_2\la 0|\big],\nn\\
&& \big(\hat{a}_1\hat{\rho}_S (t)\hat{a}^\dag_1+\hat{a}_2\hat{\rho}_S (t)\hat{a}^\dag_2\big)=e^{-\gamma t}\,|0\ra_1\la 0|\otimes|0\ra_2\la 0|,
\end{eqnarray}
and taking the time-derivative of both sides of Eq.~(\ref{N1density}), one can easily proof that Eq.~(\ref{N1density}) fulfills the Lindblad equation Eq.~(\ref{Lindblad1}).
\section{Derivation of Eqs.~(\ref{Qyfinal})}\label{G}
For notational simplicity, we define
\begin{eqnarray}
  \xi_1 &=& \alpha_1-f_1 (t),\,\,\,\,\bar{\xi}_1=\bar{\alpha}_1-\bar{f}_1 (t),\nn \\
 \eta_2 &=& \alpha_2-f_2 (t),\,\,\,\,\bar{\eta}_2=\bar{\alpha}_2-\bar{f}_2 (t),
\end{eqnarray}
then $d^2\alpha_1=d^2\xi_1$, and we have
\begin{eqnarray}
&& Q^{red}_y (\alpha_2,\bar{\alpha}_2,t)=\int d^2\alpha_1\,  Q_y (\alpha_1,\bar{\alpha}_1,\alpha_2,\bar{\alpha}_2,t),\nn\\
&& = \frac{1}{\pi^2\sigma^2_t (1-y e(t))}e^{-\frac{1}{\sigma_t}|f_2-\alpha_2|^2}\int d^2\xi_1\,e^{-\frac{1}{\sigma_t}|\xi_1|^2+\frac{y}{(1-ey)\sigma^2_t}[\xi_1 \bar{p}+\eta_2 \bar{u}][\bar{\xi}_1 p+\bar{\eta}_2 u]},\nn\\
&&= \frac{1}{\pi^2\sigma^2_t (1-y e(t))}e^{-\frac{1}{\sigma_t}|f_2-\alpha_2|^2}\,2\pi\frac{e^{\frac{y|\eta_2|^2|u|^2}{(1-ey)\sigma^2_t}}}
{\big(\frac{1}{\sigma_t}-\frac{y|p|^2}{(1-ey)\sigma^2_t}\big)}
e^{|p|^2|u|^2|\eta_2|^2\,\frac{1}{\big(\frac{1}{\sigma_t}-\frac{y|p|^2}{(1-ey)\sigma^2_t}\big)}},\nn\\
\end{eqnarray}
where in the last line we made use of the formula
\begin{equation}
  \int d^2\xi_1\,e^{-\bar{\xi}_1 z \xi_1+\bar{b}\xi_1+b\bar{\xi}_1}=\frac{2\pi}{z}\,e^{\frac{\bar{b}b}{z}},\,\,\,\Re[z]>0.
\end{equation}
Therefore,
\begin{equation}\label{QyE}
   Q^{red}_y (\alpha_2,\bar{\alpha}_2,t)=\frac{2}{\pi\,A(y)}\,e^{B(y) |\alpha_2-f_2 (t)|^2},
\end{equation}
where for convenience, we defined
\begin{eqnarray}\label{AB}
  A(y) &=& y\big(|p(t)|^2+e(t)\sigma_t\big)-\sigma_t,\nn\\
  B(y) &=& \frac{y}{(1-e(t)y)\sigma^2_t}+\frac{|p(t)|^2 (1-e(t)y)\sigma^2_t}{y|p(t)|^2-(1-e(t)y)\sigma_t}-\frac{1}{\sigma_t}.
\end{eqnarray}
\vspace{20pt}

\section*{Acknowledgements} This work has been supported by the University of Kurdistan. The authors thank Vice Chancellorship of Research and Technology, University of Kurdistan.




\end{document}